\documentclass[aps,prl,superscriptaddress,twocolumn,floatfix,longbibliography]{revtex4-2}
\usepackage{dcolumn}
\usepackage{graphicx,color}
\usepackage[utf8]{inputenc}
\usepackage{amsmath,amssymb,bm,amsfonts,dsfont,mathrsfs,amsthm}
\usepackage{braket}
\usepackage{txfonts,comment}
\usepackage[colorlinks=true,linkcolor=blue,citecolor=blue,urlcolor=blue]{hyperref}
\usepackage{soul}
\usepackage{bbm} 
\usepackage{xspace}
\usepackage{verbatim}
\usepackage{amsthm}
\usepackage{color}
\usepackage{xcolor} 

\newcommand{\red}[1]{\textcolor{black}{#1}}

\hypersetup{colorlinks=true,linkcolor=blue,citecolor=blue, filecolor=blue,urlcolor=blue,breaklinks=true}

\newtheorem{theorem}{Theorem}

\begin{document}    

\title{Overcoming Quantum Metrology Singularity through Sequential Measurements}

\author{Yaoling Yang}
\email{yyaoling@std.uestc.edu.cn}
\affiliation{Institute of Fundamental and Frontier Sciences, University of Electronic Science and Technology of China, Chengdu 611731, China}

\author{Victor Montenegro}
\email{victor.montenegro@ku.ac.ae}
\affiliation{College of Computing and Mathematical Sciences, Department of Applied Mathematics and Sciences, Khalifa University of Science and Technology, 127788 Abu Dhabi, United Arab Emirates}
\affiliation{Institute of Fundamental and Frontier Sciences, University of Electronic Science and Technology of China, Chengdu 611731, China}
\affiliation{Key Laboratory of Quantum Physics and Photonic Quantum Information, Ministry of Education, University of Electronic Science and
Technology of China, Chengdu 611731, China}

\author{Abolfazl Bayat}
\email{abolfazl.bayat@uestc.edu.cn}
\affiliation{Institute of Fundamental and Frontier Sciences, University of Electronic Science and Technology of China, Chengdu 611731, China}
\affiliation{Key Laboratory of Quantum Physics and Photonic Quantum Information, Ministry of Education, University of Electronic Science and
Technology of China, Chengdu 611731, China}

\begin{abstract}
The simultaneous estimation of multiple unknown parameters is the most general scenario in quantum sensing. Quantum multi-parameter estimation theory provides fundamental bounds on the achievable precision of simultaneous estimation. However, these bounds can become singular (no finite bound exists) in multi-parameter sensing due to parameter interdependencies, limited probe accessibility, and insufficient measurement outcomes. Here, we address the singularity issue in quantum sensing through a simple mechanism based on a sequential measurement strategy. This sensing scheme overcomes the singularity constraint and enables the simultaneous estimation of multiple parameters with a local and fixed measurement throughout the sensing protocol. This is because sequential measurements, involving consecutive steps of local measurements followed by probe evolution, inherently produce correlated measurement data that grows exponentially with the number of sequential measurements. Finally, through two different examples, namely a strongly correlated probe and a light-matter system, we demonstrate how such singularities are reflected when inferring the unknown parameters through Bayesian estimation.
\end{abstract}

\maketitle

\textit{Introduction.---} Rooted in the unique properties of quantum theory, quantum sensors have been proven to achieve higher precision compared to classical ones with the same resources~\cite{Degen,Boto2000,Leibfried-2004,Giovannetti2004, Giovannetti2006,Giovannetti2011,Demkowicz2012-nat}. In the single-parameter sensing scenario, where $\lambda$ is assumed to be the only unknown parameter to be estimated, the ultimate precision limit of $\lambda$ obeys the quantum Cram\'{e}r-Rao theorem~\cite{paris2009quantum, helstrom1969quantum,cramer1999mathematical,LeCam-1986,lehmann2006theory,Holevo}: $\mathrm{Var}(\hat{\lambda}){\geq}M^{-1}\mathcal{F}(\lambda)^{-1}{\geq}M^{-1}\mathcal{Q}(\lambda)^{-1}$. Here, $\mathrm{Var}(\hat{\lambda})$ is the variance of the local unbiased estimator of $\lambda$, $M$ is the number of experimental repetitions, and $\mathcal{F}(\lambda)$ is the classical Fisher information (CFI) with respect to $\lambda$, which quantifies the sensing precision for a fixed set of positive operator-valued measurements (POVMs) $\{\Pi_m\}$ with measurement outcome $m$~\cite{paris2009quantum,nielsen00}. By optimizing the CFI with respect to all possible POVMs one gets the quantum Fisher information (QFI) $\mathcal{Q}(\lambda){:=}\max\limits_{\{\Pi_m\}}\mathcal{F}(\lambda)$ as the ultimate sensing precision for a given quantum probe~\cite{Liu_2019}.

Although the single-parameter case is elegant and always achievable~\cite{Albarelli-2020}, practical scenarios often involve multiple unknown parameters $\bm{\lambda}{=}(\lambda_1{,}\lambda_2{,}\ldots{,}\lambda_k)$, requiring a multi-parameter sensing strategy to infer $\bm{\lambda}$~\cite{szczykulska2016multi, albarelli2023fisher,baumgratz2016quantum,humphreys2013quantum,pezze2017optimal,apellaniz2018precision,genoni2013optimal,PhysRevA.97.012106,PhysRevLett.111.070403,PhysRevA.94.042342,PhysRevA.94.062312,PhysRevLett.119.130504,vidrighin2014joint}. To address this, a natural generalization of the single-parameter results leads to the inequality~\cite{paris2009quantum, Albarelli-2020, Liu_2019,szczykulska2016multi, albarelli2023fisher}:
\begin{equation}
\bm{\mathrm{Cov}}(\bm{\hat{\lambda}}){\geq}M^{-1}\bm{\mathcal{F}}(\bm{\lambda})^{-1}{\geq}M^{-1}\bm{\mathcal{Q}}(\bm{\lambda})^{-1},\label{eq_matrix_QCRB}
\end{equation}
where $\bm{\mathrm{Cov}}(\bm{\hat{\lambda}})$ is the covariance matrix for the estimator of $\bm{\lambda}$, $\bm{\mathcal{F}}(\bm{\lambda})$ and $\bm{\mathcal{Q}}(\bm{\lambda})$ are the CFI and QFI matrices with respect to $\bm{\lambda}$, respectively. Unlike the single-parameter case, the multi-parameter scenario presents unique challenges. One such challenge is that one cannot guarantee simultaneous optimal estimation of multiple parameters due to the intrinsic non-commutativeness of quantum mechanics~\cite{ PhysRevLett.126.120503,Carollo_2019,di2022multiparameter}. Consequently, the matrix inequality $\bm{\mathcal{F}}(\bm{\lambda}){\geq}\bm{\mathcal{Q}}(\bm{\lambda})$ cannot always be saturated~\cite{szczykulska2016multi,ragy2016compatibility,PhysRevLett.126.120503,Albarelli-2020,albarelli2023fisher,Carollo_2019,paris2009quantum,Demkowicz-Dobrzański_2020,di2022multiparameter,Belliardo_2021}. 
Another issue arises when either of the Fisher information matrices becomes singular, i.e. its inverse does not exist.
This issue can occur due to several reasons: (i) parameters that seem independent in the entire quantum probe may actually be related~\cite{namkung2024unifiedcramerraoboundquantum, mihailescu2024uncertain,goldberg2021taming, Liu_2019}; (ii) even though the entire quantum probe encodes independent unknown parameters, the probe's reduced density matrix can effectively relate these independent parameters~\cite{Mihailescu_2024}; and (iii) the number of measurement outcomes might be insufficient to estimate all parameters simultaneously~\cite{Candeloro_2024}. 
Several questions may arise. First, is there any connection between the singularity of $\bm{\mathcal{F}}(\bm{\lambda})$ and $\bm{\mathcal{Q}}(\bm{\lambda})$? Second, if someone tries to blindly estimate the unknown parameters, how is the singularity of either of these matrices reflected in the final estimation? Third, how can one lift the singularity of Fisher information matrices without exploiting extra controls over the probe and measurement setup?

In this Letter, we address the above questions. First, we establish a clear connection between the singularity of the CFI and QFI matrices. Second, by considering two specific examples, we demonstrate the practical implication of singularities in blindly estimating the parameters through Bayesian estimation. Third, we show that by exploiting a sequential measurement strategy with minimal control over the system~\cite{burgarth2015quantum,montenegro2022sequential,PhysRevResearch.5.043273,bompais2022parameter,clark2019quantum,ma2018phase,benoist2019invariant,benoist2023limit,PhysRevLett.128.010604,PhysRevX.9.031009}, one can always overcome the singularity of the CFI matrix by increasing the length of the measurement sequence. This sequential measurement sensing protocol is very efficient as it satisfies the invertibility condition of the CFI matrix exponentially with respect to the measurement sequence size.

\textit{Quantum metrological tools.---} Explicit expressions of Eq.~\eqref{eq_matrix_QCRB} are now presented. The entries in the covariance matrix are: $[\mathrm{Cov}[\bm{\hat{\lambda}}]]_{ij}{=}\langle\hat{\lambda}_i\hat{\lambda}_j\rangle{-}\langle\hat{\lambda}_i\rangle\langle\hat{\lambda}_j\rangle$, where $\langle{\cdot}\rangle$ is the expected value. Matrix elements of the CFI matrix are given by:
\begin{equation}
    [\mathcal{F}(\boldsymbol{\lambda})]_{ij}{=}\sum_{m} p(m|\boldsymbol{\lambda}) \left(\partial_i \ln p(m|\boldsymbol{\lambda})\right) \left(\partial_j \ln p(m|\boldsymbol{\lambda})\right),
\end{equation}
where $\partial_j{:=}\partial/\partial\lambda_j$, $p(m|\boldsymbol{\lambda}){=}\mathrm{Tr}[\rho(\bm{\lambda})\Pi_m]$, and $\rho(\bm{\lambda})$ is the quantum state of the probe encoding a total of $k$ multiple unknown parameters $\bm{\lambda}{=}(\lambda_1{,}\lambda_2{,}\ldots{,}\lambda_k)$. The CFI matrix accounts for the simultaneous sensing precision for the multi-parameter scenario given fixed measurements basis~\cite{Albarelli-2020, Meyer_2021}. The entries of the QFI matrix, applicable to pure states as considered throughout this work, are given by:
\begin{multline}
[\mathcal{Q}(\boldsymbol{\lambda})]_{ij}{=}4\mathbb{R}e[\langle\partial_i \psi(\bm{\lambda}){|}\partial_j\psi(\bm{\lambda})\rangle\\
-\langle\partial_i\psi(\bm{\lambda}){|}\psi(\bm{\lambda})\rangle\langle\psi(\bm{\lambda}){|}\partial_j\psi(\bm{\lambda})\rangle].\label{eq_elements_QFI}
\end{multline}
The QFI matrix yields information about how well parameters can be estimated simultaneously encoded in the quantum probe~\cite{Liu_2019}. To obtain a scalar bound from the matrix inequality given in Eq.~\eqref{eq_matrix_QCRB}, it is customary to proceed as follows: multiply both sides of the inequality by a positive, real weighting matrix $\mathcal{W}$ of dimension $k{\times}k$, and then take the trace of the resulting expressions. This process gives the scalar inequality: $\mathrm{Tr}[\mathcal{W}\bm{\mathrm{Cov}}(\bm{\hat{\lambda}})]{\geq}M^{-1}\mathrm{Tr}[\mathcal{W}\bm{\mathcal{F}}(\bm{\lambda})^{-1}]{\geq}M^{-1}\mathrm{Tr}[\mathcal{W}\bm{\mathcal{Q}}(\bm{\lambda})^{-1}]$. The choice of $\mathcal{W}$ can prioritize the uncertainty of certain parameters over others. Without loss of generality, we consider $\mathcal{W}{=}\mathbbm{1}_{k{\times}k}$. Thus, the scalar bound reads as:
\begin{equation}
    \mathrm{Tr}[\bm{\mathrm{Cov}}(\bm{\hat{\lambda}})]{\geq}M^{-1}\mathrm{Tr}[\bm{\mathcal{F}}(\bm{\lambda})^{-1}]{\geq}M^{-1}\mathrm{Tr}[\bm{\mathcal{Q}}(\bm{\lambda})^{-1}],\label{eq_trace_QCRB}
\end{equation}
which gives equal importance to all unknown parameters as $\mathrm{Tr}[\bm{\mathrm{Cov}}(\bm{\hat{\lambda}})]{=}\sum_{j=1}^k\langle\lambda_j^2\rangle{-}\langle\lambda_j\rangle^2{=}\sum_{j=1}^k\mathrm{Var}(\lambda_j)$.

\textit{Singularity of Fisher information matrices.---} The bounds on the covariance matrix in the Cram\'{e}r-Rao inequality, as shown in Eq.~(\ref{eq_matrix_QCRB}), are determined by the inverse of Fisher information matrices. However, if these matrices are singular, the corresponding bounds become undefined, resulting in arbitrarily large uncertainties in parameter estimation. A singularity can be resolved using either the Moore-Penrose pseudoinverse of the CFI matrix~\cite{namkung2024unifiedcramerraoboundquantum} or, alternatively, by deriving the CRB as the solution to an unconstrained quadratic maximization problem~\cite{stoica2001parameter}. In general, this singularity poses a fundamental challenge for sensing applications. In this section, we examine the relationship between the singularity of the CFI and the QFI. 

\begin{theorem}
The singularity of the QFI matrix $\bm{\mathcal{Q}}(\bm{\lambda})$ implies the singularity of the CFI matrix $\bm{\mathcal{F}}(\bm{\lambda})$ for any chosen measurement basis.\label{theorem1}
\end{theorem}
The proof of the theorem can be found in the Supplemental Material (SM)~\cite{SM_multiparameter}. \red{Note that if the matrix $\bm{\mathcal{Q}}(\bm{\lambda})$ is singular, then the tighter Holevo Cram\'{e}r-Rao bound $C_{\boldsymbol{\lambda}}^{H}$ also becomes ill-defined. This is evident from the inequality $C_{\boldsymbol{\lambda}}^{H}{\geq}\max\left\{\textrm{Tr}\left[W (\mathcal{Q}^{-1})\right],\, C_{\boldsymbol{\lambda}}^{RLD}\right\}$, where $C_{\boldsymbol{\lambda}}^{RLD}$ is the right logarithmic derivative (RLD) bound~\cite{holevo2011probabilistic,albarelli2019evaluating}. Since the lower bound of the above Holevo Cram\'{e}r-Rao inequality involves the inverse of the matrix $\mathcal{Q}$, a singular $\mathcal{Q}$ would still render the Holevo bound $C_{\boldsymbol{\lambda}}^{H}$ to be ill-defined.}
Although the singularity of the QFI matrix $\bm{\mathcal{Q}}(\bm{\lambda})$ directly implies the singularity of the CFI matrix $\bm{\mathcal{F}}(\bm{\lambda})$ for any measurement basis, the reverse connection is less straightforward. \red{Indeed, the QFI matrix singularity reflects a limitation of the probe state itself, whereas the singularity of the CFI matrix indicates that the chosen measurement basis fails to extract sufficient information.}

To exemplify this, let us consider two unknown parameters $\bm{\lambda}{=}(\theta{,} \phi)$ to be estimated simultaneously encoded in a qubit $\ket{\psi}{=}\cos\frac{\theta}{2}\ket{\uparrow}{+}e^{i\phi}\sin\frac{\theta}{2}\ket{\downarrow}$. One can evaluate the elements of the QFI matrix using Eq.~\eqref{eq_elements_QFI}, that is: $[\mathcal{Q}(\bm{\lambda})]_{\theta\theta}{=}1$, $[\mathcal{Q}(\bm{\lambda})]_{\phi\phi}{=}\sin^2 \theta$, and $[\mathcal{Q}(\bm{\lambda})]_{\theta\phi}{=}[\mathcal{Q}(\bm{\lambda})]_{\phi\theta}{=}0$. These result in the invertibility of the QFI matrix, implying that the simultaneous estimation of $(\theta{,}\phi)$ is indeed possible. Nonetheless, any two-outcome projective measurement basis $|\Upsilon_1\rangle{=}\cos\frac{\theta'}{2}\ket{\uparrow}{+}e^{i\phi'}\sin\frac{\theta'}{2}\ket{\downarrow}$ and $|\Upsilon_2\rangle{=}\sin\frac{\theta'}{2}\ket{\uparrow}{-}e^{i\phi'}\cos\frac{\theta'}{2}\ket{\downarrow}$, each appearing with probability $p(m|\bm{\lambda}){=}|\langle\Upsilon_m|\psi\rangle|^2$ (for $m{=}1,2$), results in a singular CFI regardless of the choice of $(\theta'{,}\phi')$, see the SM~\cite{SM_multiparameter} for detailed calculations, \red{including a straightforward generalization of the results to the case of a mixed qubit state.} This presents an intriguing scenario: the non-singularity of the QFI matrix confirms that parameter estimation is indeed possible. However, the singularity of the CFI matrix indicates that no two-outcome projective measurement can accomplish the estimation task. It is important to note that the impossibility of constructing an estimator $\bm{\hat{\lambda}}$ that maps two measurement outcomes to the two-dimensional parameter space arises because no continuous bijection exists between $p(1|\bm{\lambda}){\in}\mathbb{R}^1$ [with $p(2|\bm{\lambda}){=}1{-}p(1|\bm{\lambda})$] and $\bm{\lambda}{=}(\theta{,}\phi){\in}\mathbb{R}^2$~\cite{munkres2018elements}. To address this issue, one can resort to POVMs, where the orthogonality between the set of measurement operators is relaxed, allowing for an increased number of measurement outcomes. See the SM for details~\cite{SM_multiparameter}. This leads us to a theorem on the singularity of the CFI matrix and its connection to the space of probability distributions:
\begin{theorem}
For a random variable $X$ with $m$ outcomes described by the probability distribution $\{p(i|\boldsymbol{\lambda})\}_{i=1}^m$, which depends on $k$ unknown parameters $\boldsymbol{\lambda}{=}(\lambda_1{,}\ldots{,}\lambda_k)$, the corresponding CFI matrix is not invertible if $m{<}k{+}1$.\label{theorem2}
\end{theorem}
The proof of the theorem is provided in Ref.~\cite{Candeloro_2024}. As stated in the theorem, a practical approach to resolve the singularity of the CFI matrix is to increase the number of measurement outcomes $m$, ensuring that $m{\geq}k{+}1$. This condition allows for the simultaneous estimation of $k$ unknown parameters. In the following, we introduce a protocol that utilizes a sequence of single-qubit measurements in a fixed projective basis. This approach exponentially increases the number of outcomes $m$, inherently satisfying the condition $m{\geq}k{+}1$ required for the simultaneous estimation of $k$ unknown parameters.

\textit{Sequential measurement sensing.---} 
The sequential measurement sensing protocol~\cite{burgarth2015quantum,bompais2022parameter,montenegro2022sequential,PhysRevResearch.5.043273,yang2024sequential} proceeds as: (i) prepare an initial quantum state; (ii) let the state evolve for a time $\tau$, during which it encodes $k$ unknown parameters; (iii) measure a part of the system (i.e., a local measurement) in a fixed basis and record the measurement outcome; (iv) without resetting the system repeat steps (ii) and (iii) for $n_\mathrm{seq}$ iterations to collect a sequence of measurement outcomes. After completing $n_\mathrm{seq}$ sequential measurements, the quantum probe is reset to its initial state, allowing a fresh sequence of measurement outcomes to be collected. This sensing protocol naturally creates correlated data, with a total measurement outcome space of $m{=}2^{n_\mathrm{seq}}$. Thus, according to Theorem~\ref{theorem2}, the condition $2^{n_\mathrm{seq}}{\geq}k{+}1$ allows for multi-parameter sensing using very few sequential projective measurements in a given basis. \red{Note that the choice of $\tau$ must be long enough to allow quantum correlations to spread across the system, in accordance with the Lieb-Robinson bound~\cite{lieb1972finite}. If $\tau$ is too small, it can suppress dynamics due to the quantum Zeno effect~\cite{facchi2008quantum}. On the other hand, if $\tau$ is too large, it wastes total protocol time~\cite{montenegro2022sequential}. For simplicity, we assume that $\tau$ scales with the system size. Further optimizations are discussed in the SM~\cite{SM_multiparameter}.} In the following, we provide two distinct probes to show the sequential measurements for multi-parameter sensing.

\textit{Model I: a spin chain probe.---} Quantum many-body probes are known to exhibit quantum-enhanced sensitivity in both equilibrium~\cite{Ding2022, Liu2021, zanardi2008quantum, invernizzi2008optimal, salvatori2014quantum, zanardi2007critical, Garbe2020, Jianming2021critical, Horodecki2018prx, Sarkar2022topological, PhysRevX.8.021022, Montenegro2021} and non-equilibrium~\cite{fernandez2017quantum,baumann2010dicke,baden2014realization,klinder2015dynamical,rodriguez2017probing,fitzpatrick2017observation,fink2017observation,ilias2022criticality,Ilias2023Criticality,Alipour2014Quantum, mishra2021driving,mishra2022integrable, montenegro2023quantum,Cabot2024Continuous,lyu2020eternal,Iemini2023,yousefjani2024discrete}. Specifically, we consider a one-dimensional spin chain composed of $N$ interacting spin-$1/2$ particles with Heisenberg interaction:
\begin{equation}
H = -J \sum_{j=1}^{N-1} \boldsymbol{\sigma}^j \cdot \boldsymbol{\sigma}^{j+1} + \sum_{j=1}^{k}B_x^{(j)}\sigma_x^j,\label{eq_heisenberg_model}
\end{equation}
where $J$ is the exchange interaction between particles and $\boldsymbol{\sigma}^j$ is a vector of Pauli matrices $\boldsymbol{\sigma}^j{=}(\sigma_x^j{,}\sigma_y^j{,}\sigma_z^j)$ at site $j$. We aim to estimate $k$ unknown magnetic fields $\bm{B}{=}(B_x^{(1)}{,}B_x^{(2)}{,}\ldots{,}B_x^{(k)})$ along the $x-$direction with $k{\leq}N$. We measure only a single spin at site $N$ sequentially in a fixed $\sigma_z$ basis; $\sigma_z^j\ket{\uparrow}{=}\ket{\uparrow}$ and $\sigma_z^j\ket{\downarrow}{=}{-}\ket{\downarrow}$. Without loss of generality, the time interval between consecutive measurements is set to $J\tau{=}N$, and the initial state is chosen as $\ket{\downarrow\downarrow\ldots\downarrow}$. 

\textit{Model II: a light-matter probe.---} Light-matter probes are a mature platform for quantum sensing and metrology~\cite{kucsko2013nanometre,taylor2008high,garbe2020critical,Barrett_2016,karnieli2023quantum}. We consider two non-interacting two-level atoms coupled to a single-mode quantized electromagnetic field via Jaynes-Cummings Hamiltonian:
\begin{equation}
H_\mathrm{JC} = \omega_a a^\dagger a + \sum_{j=1}^{2} \frac{\omega_j}{2} \sigma_{z}^j + \sum_{j=1}^{2} J_j \left( a^\dagger \sigma_{-}^j + a \sigma_{+}^j \right),\label{eq_JC_model}
\end{equation}
where $a$ ($a^\dagger$) is the annihilation (creation) operator for the quantized field obeying $[a{,}a^\dagger]{=}1$, and  $\sigma_+^j{=}\ket{\uparrow}\bra{\downarrow}$ ($\sigma_-^j{=}\ket{\downarrow}\bra{\uparrow}$) is the raising (lowering) spin operator at site $j$. The $j$-th two-level atom of frequency $\omega_j$ couples to the field mode of frequency $\omega_a$ with an interaction strength denoted by $J_j$. We aim to estimate the parameters $\omega_j$ and $J_j$ by sequentially measuring a single two-level atom in $\sigma_z$ basis. Without loss of generality, we fix the measurements time interval as $\omega_a \tau{=}2\pi$.

\textit{Overcoming the CFI matrix singularity.---} 
To show that sequential measurements are a natural strategy to overcome the singularity of the CFI matrix, we first evaluate the trace of the inverse of the CFI matrix in Eq.~\eqref{eq_trace_QCRB} for the Heisenberg model described in Eq.~\eqref{eq_heisenberg_model}. Consequently, we investigate the singularity of the CFI matrix in the presence of $k$ multiple unknown magnetic fields $\bm{B}{=}(B_x^{(1)}{,}B_x^{(2)}{,}\ldots{,}B_x^{(k)})$, namely $\mathrm{Tr}[\mathcal{F}(\bm{B})^{-1}]$.
\begin{figure}[t]
\includegraphics[width=\linewidth]{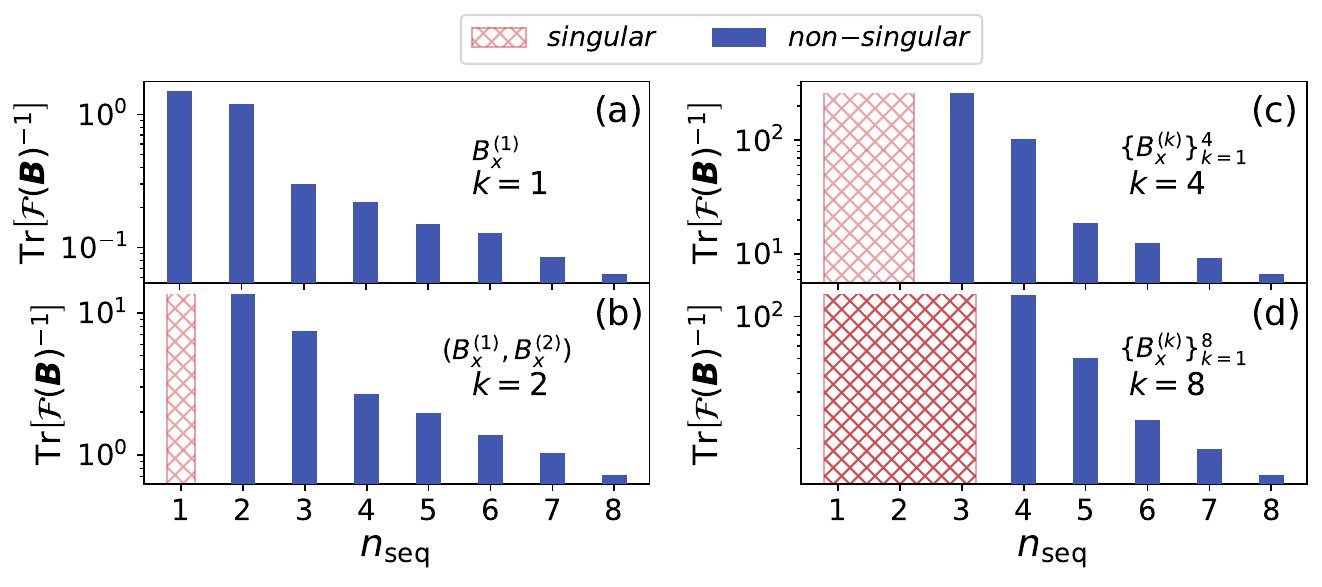} 
\caption{Heisenberg probe: Trace of the inverse of the CFI matrix $\mathrm{Tr}[\mathcal{F}(\bm{B})^{-1}]$ as a function of the number of sequential measurements $n_\mathrm{seq}$ for different $k$ unknown magnetic fields $\bm{B}{=}(B_x^{(1)}{,}B_x^{(2)}{,}\ldots{,}B_x^{(k)})$. Blue bars (red cross-hatched) represent non-singular (singular) CFI matrix. We consider $N{=}10$ and $B_x^{(k)}{=}0.5J$ $\forall k$.}\label{fig_trace_inverse_FIM} 
\end{figure}
In Fig.~\ref{fig_trace_inverse_FIM}, we plot the trace of the inverse of the CFI matrix $\mathrm{Tr}[\mathcal{F}(\bm{B})^{-1}]$ as a function of the number of sequential measurements $n_\mathrm{seq}$ for different unknown magnetic fields $\{B_x^{(k)}\}$. Red cross-hatched regions indicate areas where the corresponding CFI matrix is singular (non-invertible), thus it is not possible to estimate the $k$ unknown magnetic fields. In contrast, blue bars represent regions where the CFI matrix is non-singular (invertible), hence the estimation of the $k$ unknown magnetic fields is possible. In Fig.~\ref{fig_trace_inverse_FIM}(a), we show the case where $k{=}1$, i.e., single-parameter sensing. In this scenario, it is always possible to estimate a single unknown magnetic field because there are always at least two measurement outcomes available. Consequently, the condition $2^{n_\mathrm{seq}}{\geq}k{+}1$ is satisfied for any $n_\mathrm{seq}{\geq}1$. In Figs.~\ref{fig_trace_inverse_FIM}(b)-(d), by increasing the number of unknown parameters $k$ two evident features emerge. First, a clear transition from singularity to non-singularity is observed, occurring precisely when the condition $2^{n_\mathrm{seq}}{\geq}k{+}1$ is met. Specifically, to estimate $k{=}2{,}4{,}8$ unknown parameters, at least $n_\mathrm{seq}{=}2{,}3{,}4$ sequential measurements are required, respectively. Second, $\mathrm{Tr}[\mathcal{F}(\bm{B})^{-1}]$ decreases as the number of sequential measurements $n_\mathrm{seq}$ increases, implying that the uncertainty in estimating $\bm{B}$ decreases.
\begin{figure}[t]
\includegraphics[width=\linewidth]{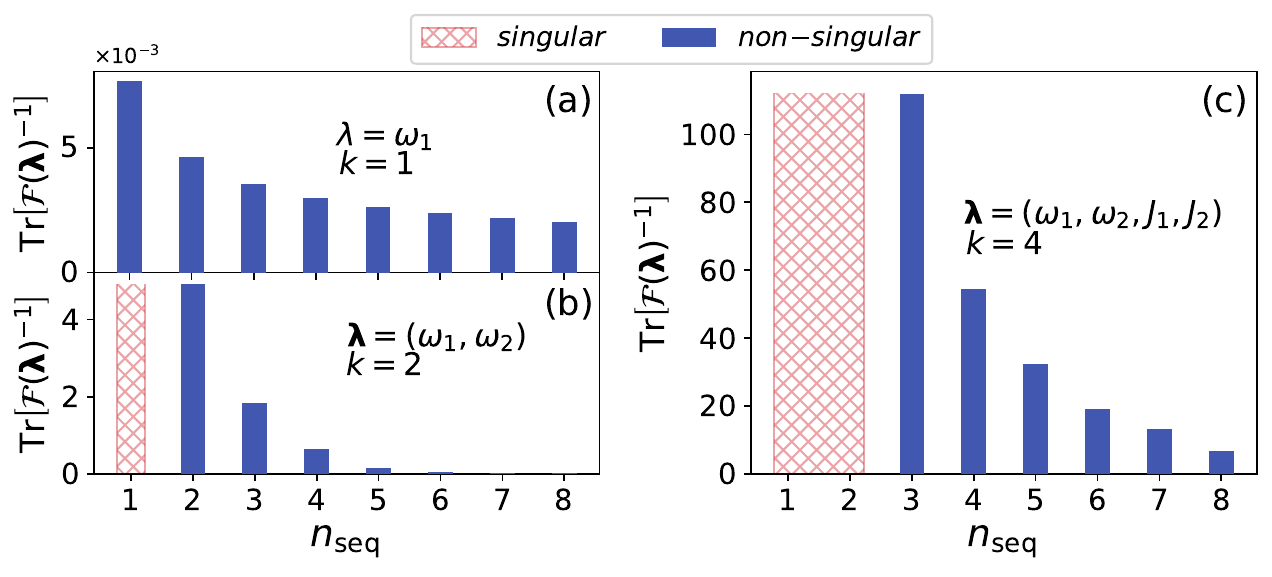} 
\caption{The light-matter probe: Trace of the inverse of the CFI matrix $\mathrm{Tr}[\mathcal{F}(\bm{\lambda})^{-1}]$ as a function of the number of sequential measurements $n_\mathrm{seq}$ for different $k$ unknown parameters $\bm{\lambda}{=}(\omega_1{,}\omega_2{,}J_1{,}J_2)$. Blue bars (red cross-hatched) represent non-singular (singular) CFI matrix. We consider $\bm{\lambda}{=}(0.9{,}1.1{,}0.1{,}0.2)\omega_a$ and $|\psi(0)\rangle{=}|\downarrow{,}\downarrow{,}\alpha\rangle$ with $\alpha{=}2$.} \label{fig_JC} 
\end{figure}

Similar conclusions can be drawn for the light-matter probe described by Eq.~\eqref{eq_JC_model}. In Fig.~\ref{fig_JC}, we plot the trace of the inverse of the CFI matrix $\text{Tr}[\mathcal{F}(\bm{\lambda})^{-1}]$ as a function of the number of sequential measurements $n_\mathrm{seq}$ for different $k$ multiple unknown parameters $\bm{\lambda}{=}(\omega_1{,}\omega_2{,}J_1{,}J_2)$. The figure shows that the singularity is resolved precisely when the condition $ 2^{n_\mathrm{seq}}{\geq}k{+}1$ is met. Furthermore, as the number of sequential measurements $n_\mathrm{seq}$ increases, the uncertainty in estimating multiple simultaneous parameters systematically decreases. Thus, as demonstrated in both examples, the sequential measurement strategy efficiently overcomes the singularity of the CFI matrix while reducing the uncertainty in estimating multiple parameters simultaneously. Notably, the proposed strategy operates with minimal control over the probe, local access to the probe, and a single projective measurement throughout the entire sensing process.

\textit{Bayesian estimation.---}
The inverse of the Fisher information matrices in Eq.~(\ref{eq_matrix_QCRB}) provides only a lower bound on the covariance matrix of the estimators. This raises an important question: What is the actual implication of the singularity of these matrices on the ability to estimate the unknown parameters? To answer this, we directly estimate the unknown parameters from measurement outcomes using a Bayesian approach, which is an optimal estimator in the asymptotic limit of measurement data~\cite{paris2009quantum,cramer1999mathematical,LeCam-1986,lehmann2006theory}. We apply this approach blindly, without considering whether the CFI matrix is singular. The Bayesian estimator relies on Bayes' theorem:
\begin{equation}
    P(\bm{\lambda}|\text{data}) = \frac{P(\text{data}|\bm{\lambda})P(\bm{\lambda})}{P(\text{data})},\label{eq_Bayes_rule}
\end{equation}
where $P(\bm{\lambda}|\text{data})$ is the \textit{posterior}, representing the probability distribution for $\bm{\lambda}$ given the measurement data. $P(\text{data}|\bm{\lambda})$, known as the \textit{likelihood}, is the probability distribution for the observed measurement data assuming $\bm{\lambda}$ is known. $P(\bm{\lambda})$ is the \textit{prior} information available, and $P(\text{data})$ is a normalization constant that ensures the posterior behaves as a probability distribution~\cite{Borovkov1984,vantrees1968}. For the sake of simplicity and without loss of generality we assume discrete parameter space and uniform distribution for the prior over a finite interval.  

\begin{figure}[t]
\includegraphics[width=\linewidth]{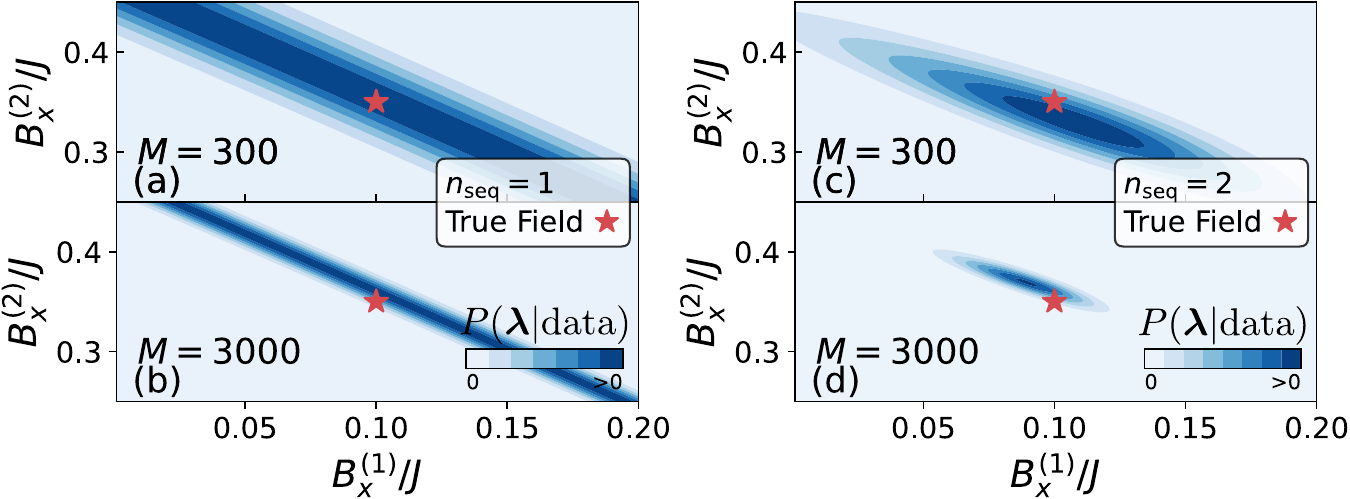}  
\caption{Posterior distributions $P(\bm{\lambda}|\mathrm{data})$ as functions of $B_x^{(1)}$ and $B_x^{(2)}$ for different $n_\mathrm{seq}$ and $M$. Left (right) column corresponds to $n_\mathrm{seq}{=}1$ ($n_\mathrm{seq}{=}2$). The true field is $(B_x^{(1)}{,}B_x^{(2)}){=}(0.1{,}0.35)J$. The system size is $N=6$.} \label{fig_posterior} 
\end{figure}

We first consider the Heisenberg probe in Eq.~\eqref{eq_heisenberg_model} with two unknown parameters $\bm{B}{=}(B_x^{(1)}{,}B_x^{(2)})$. In Figs.~\ref{fig_posterior}(a)-(b) we plot the posterior for the case of $n_\mathrm{seq}{=}1$, where the CFI matrix is singular, for two different samples $M$. 
As the figures clearly show, the posterior distribution does not converge to the true values, regardless of $M$. In fact, increasing $M$ only narrows the strip-shaped posterior, but it remains extended in the parameter space, indicating that the estimation procedure cannot distinguish which are the actual unknown parameters. Thus, significant uncertainty about the parameters persists for any $M$. In contrast, in Figs.~\ref{fig_posterior}(c)-(d), simultaneous estimation of $B_x^{(1)}$ and $B_x^{(2)}$ becomes possible by increasing the number of sequential measurements to $n_\mathrm{seq}{=}2$. This occurs because the CFI matrix is non-singular, as $2^2{\geq}2{+}1$. Furthermore, increasing $M$ leads to more accurate estimation of the parameters. A similar analysis leading to the same conclusion can be done for the light-matter model of Eq.~\eqref{eq_JC_model}, see SM~\cite{SM_multiparameter} for details.

\begin{figure}[t]
\includegraphics[width=\linewidth]{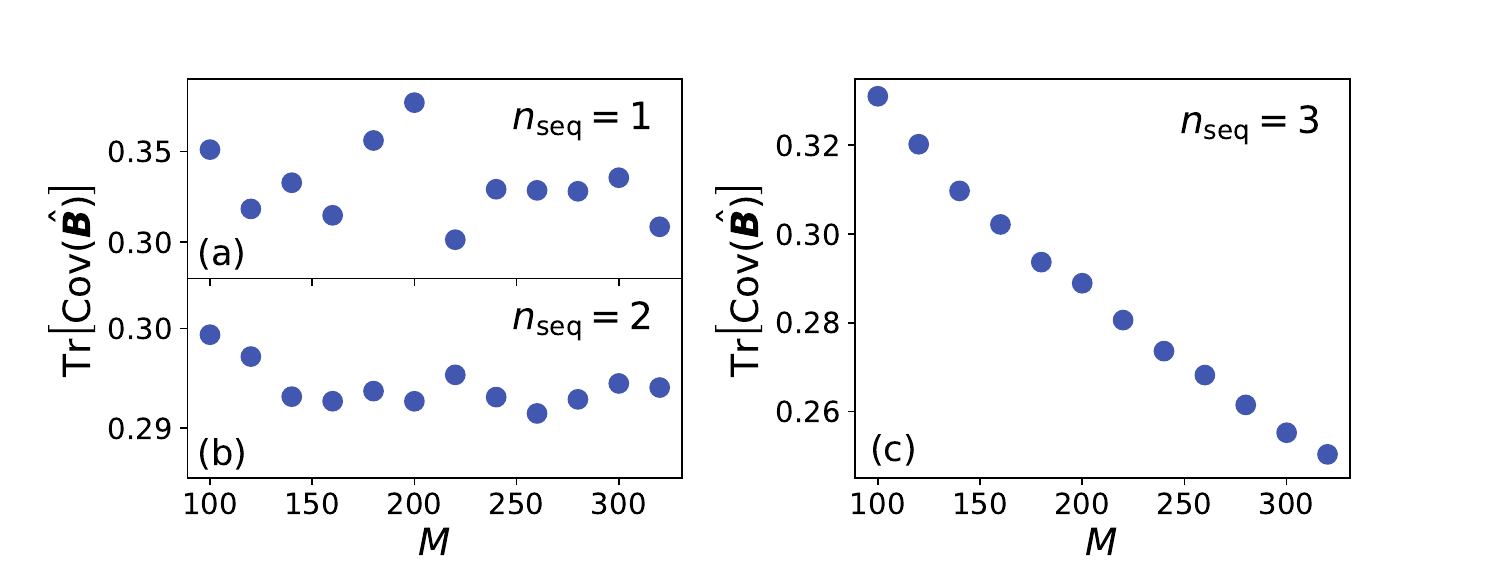} 
\caption{Trace of covariance matrix $\text{Tr}(\text{Cov}[\hat{\boldsymbol{B}}])$ using a Bayesian estimator as a function of $M$. We encode $k{=}4$ unknown magnetic fields $\bm{B}{=}(B_x^{(1)}{,}B_x^{(2)}{,}B_x^{(3)}{,}B_x^{(4})$ where $B_x^{(k)}{=}0.4J$ for $1{\leq}k{\leq}4$. (a) $n_\text{seq}{=}1$, (b) $n_\text{seq}{=}2$, and (c) $n_\text{seq}{=} 3$. The system size is set to $N=6$ and $\mu = 10^4$.}\label{fig_trace_cov} 
\end{figure}

To further investigate the estimability of $\bm{B}$, we analyze the covariance matrix of the estimator $\hat{\bm{B}}$ in Eq.~\eqref{eq_trace_QCRB}. To do so, we determine the estimated field $\bm{B}_\mathrm{est}$ as the coordinates of the unknown true field $\bm{B}$ that maximize the posterior function for a given $M$:
\begin{equation}
    \boldsymbol{B}_{\mathrm{est},i} = \underset{\bm{B}}{\arg\max} \, P(\boldsymbol{B}|\text{data}),
\end{equation}
where the label $\mathrm{est},i$ denotes the estimation of $\bm{B}$ for the $i$th sample. By generating $\mu$ sets of samples (each $\mu$ returns a single estimate of $\bm{B}$ from a total of $M$ measurements), we can construct the vector of estimated values $(\boldsymbol{B}_{\mathrm{est},1}{,}\boldsymbol{B}_{\mathrm{est},2}{,}\ldots{,}\boldsymbol{B}_{\mathrm{est},\mu})$ and straightforwardly obtain the matrix $\mathrm{Cov}[\hat{\boldsymbol{B}}]$ averaging over $\mu$. In Fig.~\ref{fig_trace_cov}, we plot the trace of the covariance matrix $\text{Tr}(\text{Cov}[\hat{\bm{B}}])$ as a function of the number of $M$ for different $n_\mathrm{seq}$; we encode four parameters $\{B_x^{(k)}\}_{k=1}^4$. Figs.~\ref{fig_trace_cov}(a)-(b) show that $n_\mathrm{seq}{<}3$ is insufficient to estimate $k{=}4$ parameters, as $\text{Tr}(\text{Cov}[\hat{\boldsymbol{B}}])$ remains nearly constant as $M$ increases. This aligns with the singularity of the CFI matrix for $2^{n_\mathrm{seq}}{<}k{+}1$. In contrast, Fig.~\ref{fig_trace_cov}(c) shows that for $n_\mathrm{seq}{=}3$, $\text{Tr}(\text{Cov}[\hat{\bm{B}}])$ clearly decreases with the number of measurement repetitions $M$. In agreement with the non-singular condition $2^{n_\mathrm{seq}}{\geq}k{+}1$. This demonstrates the effectiveness of simultaneously estimating multiple parameters using sequential measurements. Same conclusion can be found for the light-matter model of Eq.~\eqref{eq_JC_model}, see SM~\cite{SM_multiparameter} for details.

\textit{Protocol optimization.---} \red{So far, our sequential measurement protocol has used a fixed time interval between consecutive measurements and a fixed measurement basis. Note that although optimization of such parameters cannot avoid the singularity issue, it can improve the precision once the measurement sequence size is large enough and singularity issue is resolved. Indeed, such optimization leads to a substantial reduction in the estimation uncertainty $\mathrm{Tr}[\mathcal{F}(\bm{\lambda})^{-1}]$---see the SM~\cite{SM_multiparameter} for details.}

\textit{Conclusions.---} In this Letter, we address a key challenge in multi-parameter quantum sensing: the singularity (non-invertibility) of Fisher information matrices. \red{We begin by establishing the connection between the singularity of the CFI and QFI matrices. The central problem we address is the scenario in which the QFI matrix is non-singular, while the corresponding CFI matrix remains singular due to insufficiently informative measurement outcomes.} We then introduce a systematic approach to overcome the singularity of the CFI matrix through a sequential measurement sensing strategy that harnesses temporal correlations via measurement-induced wavefunction collapse. This approach exploits the exponential growth of the probability space with the number of sequential measurements. As a result, our sensing protocol---applicable to general quantum systems---enables the efficient simultaneous estimation of multiple unknown parameters. \red{While our presented proposal operates under minimal control assumptions, we show that further optimizations, such as adapting both the measurement basis and evolution time between consecutive measurements, can lead to reduced uncertainties.} Finally, we demonstrate that postprocessing the measurement data with a Bayesian estimator reveals the impact of singularities on the estimation procedure and confirms that our sequential measurement scheme effectively overcomes these challenges in two distinct quantum probes.

\emph{Acknowledgments.---} AB acknowledges support from National Natural Science Foundation of China (Grants Nos. 12050410253, 92065115, and 12274059) and the Ministry of Science and Technology of China (Grant No. QNJ2021167001L). VM thanks support from the National Natural Science Foundation of China (Grants No. 12374482 and No. W2432005).

\bibliographystyle{apsrev4-2}

\bibliography{multiparams}

\begin{thebibliography}{95}%
\makeatletter
\providecommand \@ifxundefined [1]{%
 \@ifx{#1\undefined}
}%
\providecommand \@ifnum [1]{%
 \ifnum #1\expandafter \@firstoftwo
 \else \expandafter \@secondoftwo
 \fi
}%
\providecommand \@ifx [1]{%
 \ifx #1\expandafter \@firstoftwo
 \else \expandafter \@secondoftwo
 \fi
}%
\providecommand \natexlab [1]{#1}%
\providecommand \enquote  [1]{``#1''}%
\providecommand \bibnamefont  [1]{#1}%
\providecommand \bibfnamefont [1]{#1}%
\providecommand \citenamefont [1]{#1}%
\providecommand \href@noop [0]{\@secondoftwo}%
\providecommand \href [0]{\begingroup \@sanitize@url \@href}%
\providecommand \@href[1]{\@@startlink{#1}\@@href}%
\providecommand \@@href[1]{\endgroup#1\@@endlink}%
\providecommand \@sanitize@url [0]{\catcode `\\12\catcode `\$12\catcode
  `\&12\catcode `\#12\catcode `\^12\catcode `\_12\catcode `\%12\relax}%
\providecommand \@@startlink[1]{}%
\providecommand \@@endlink[0]{}%
\providecommand \url  [0]{\begingroup\@sanitize@url \@url }%
\providecommand \@url [1]{\endgroup\@href {#1}{\urlprefix }}%
\providecommand \urlprefix  [0]{URL }%
\providecommand \Eprint [0]{\href }%
\providecommand \doibase [0]{https://doi.org/}%
\providecommand \selectlanguage [0]{\@gobble}%
\providecommand \bibinfo  [0]{\@secondoftwo}%
\providecommand \bibfield  [0]{\@secondoftwo}%
\providecommand \translation [1]{[#1]}%
\providecommand \BibitemOpen [0]{}%
\providecommand \bibitemStop [0]{}%
\providecommand \bibitemNoStop [0]{.\EOS\space}%
\providecommand \EOS [0]{\spacefactor3000\relax}%
\providecommand \BibitemShut  [1]{\csname bibitem#1\endcsname}%
\let\auto@bib@innerbib\@empty
\bibitem [{\citenamefont {Degen}\ \emph {et~al.}(2017)\citenamefont {Degen},
  \citenamefont {Reinhard},\ and\ \citenamefont {Cappellaro}}]{Degen}%
  \BibitemOpen
  \bibfield  {author} {\bibinfo {author} {\bibfnamefont {C.~L.}\ \bibnamefont
  {Degen}}, \bibinfo {author} {\bibfnamefont {F.}~\bibnamefont {Reinhard}},\
  and\ \bibinfo {author} {\bibfnamefont {P.}~\bibnamefont {Cappellaro}},\
  }\href {https://doi.org/10.1103/RevModPhys.89.035002} {\bibfield  {journal}
  {\bibinfo  {journal} {Rev. Mod. Phys.}\ }\textbf {\bibinfo {volume} {89}},\
  \bibinfo {pages} {035002} (\bibinfo {year} {2017})}\BibitemShut {NoStop}%
\bibitem [{\citenamefont {Boto}\ \emph {et~al.}(2000)\citenamefont {Boto},
  \citenamefont {Kok}, \citenamefont {Abrams}, \citenamefont {Braunstein},
  \citenamefont {Williams},\ and\ \citenamefont {Dowling}}]{Boto2000}%
  \BibitemOpen
  \bibfield  {author} {\bibinfo {author} {\bibfnamefont {A.~N.}\ \bibnamefont
  {Boto}}, \bibinfo {author} {\bibfnamefont {P.}~\bibnamefont {Kok}}, \bibinfo
  {author} {\bibfnamefont {D.~S.}\ \bibnamefont {Abrams}}, \bibinfo {author}
  {\bibfnamefont {S.~L.}\ \bibnamefont {Braunstein}}, \bibinfo {author}
  {\bibfnamefont {C.~P.}\ \bibnamefont {Williams}},\ and\ \bibinfo {author}
  {\bibfnamefont {J.~P.}\ \bibnamefont {Dowling}},\ }\href
  {https://doi.org/10.1103/PhysRevLett.85.2733} {\bibfield  {journal} {\bibinfo
   {journal} {Phys. Rev. Lett.}\ }\textbf {\bibinfo {volume} {85}},\ \bibinfo
  {pages} {2733} (\bibinfo {year} {2000})}\BibitemShut {NoStop}%
\bibitem [{\citenamefont {Leibfried}\ \emph {et~al.}(2004)\citenamefont
  {Leibfried}, \citenamefont {Barrett}, \citenamefont {Schaetz}, \citenamefont
  {Britton}, \citenamefont {Chiaverini}, \citenamefont {Itano}, \citenamefont
  {Jost}, \citenamefont {Langer},\ and\ \citenamefont
  {Wineland}}]{Leibfried-2004}%
  \BibitemOpen
  \bibfield  {author} {\bibinfo {author} {\bibfnamefont {D.}~\bibnamefont
  {Leibfried}}, \bibinfo {author} {\bibfnamefont {M.~D.}\ \bibnamefont
  {Barrett}}, \bibinfo {author} {\bibfnamefont {T.}~\bibnamefont {Schaetz}},
  \bibinfo {author} {\bibfnamefont {J.}~\bibnamefont {Britton}}, \bibinfo
  {author} {\bibfnamefont {J.}~\bibnamefont {Chiaverini}}, \bibinfo {author}
  {\bibfnamefont {W.~M.}\ \bibnamefont {Itano}}, \bibinfo {author}
  {\bibfnamefont {J.~D.}\ \bibnamefont {Jost}}, \bibinfo {author}
  {\bibfnamefont {C.}~\bibnamefont {Langer}},\ and\ \bibinfo {author}
  {\bibfnamefont {D.~J.}\ \bibnamefont {Wineland}},\ }\href
  {https://doi.org/10.1126/science.1097576} {\bibfield  {journal} {\bibinfo
  {journal} {Science}\ }\textbf {\bibinfo {volume} {304}},\ \bibinfo {pages}
  {1476} (\bibinfo {year} {2004})}\BibitemShut {NoStop}%
\bibitem [{\citenamefont {Giovannetti}\ \emph {et~al.}(2004)\citenamefont
  {Giovannetti}, \citenamefont {Lloyd},\ and\ \citenamefont
  {Maccone}}]{Giovannetti2004}%
  \BibitemOpen
  \bibfield  {author} {\bibinfo {author} {\bibfnamefont {V.}~\bibnamefont
  {Giovannetti}}, \bibinfo {author} {\bibfnamefont {S.}~\bibnamefont {Lloyd}},\
  and\ \bibinfo {author} {\bibfnamefont {L.}~\bibnamefont {Maccone}},\
  }\href@noop {} {\bibfield  {journal} {\bibinfo  {journal} {Science}\ }\textbf
  {\bibinfo {volume} {306}},\ \bibinfo {pages} {1330} (\bibinfo {year}
  {2004})}\BibitemShut {NoStop}%
\bibitem [{\citenamefont {Giovannetti}\ \emph {et~al.}(2006)\citenamefont
  {Giovannetti}, \citenamefont {Lloyd},\ and\ \citenamefont
  {Maccone}}]{Giovannetti2006}%
  \BibitemOpen
  \bibfield  {author} {\bibinfo {author} {\bibfnamefont {V.}~\bibnamefont
  {Giovannetti}}, \bibinfo {author} {\bibfnamefont {S.}~\bibnamefont {Lloyd}},\
  and\ \bibinfo {author} {\bibfnamefont {L.}~\bibnamefont {Maccone}},\
  }\href@noop {} {\bibfield  {journal} {\bibinfo  {journal} {Phys. Rev. Lett.}\
  }\textbf {\bibinfo {volume} {96}},\ \bibinfo {pages} {010401} (\bibinfo
  {year} {2006})}\BibitemShut {NoStop}%
\bibitem [{\citenamefont {Giovannetti}\ \emph {et~al.}(2011)\citenamefont
  {Giovannetti}, \citenamefont {Lloyd},\ and\ \citenamefont
  {Maccone}}]{Giovannetti2011}%
  \BibitemOpen
  \bibfield  {author} {\bibinfo {author} {\bibfnamefont {V.}~\bibnamefont
  {Giovannetti}}, \bibinfo {author} {\bibfnamefont {S.}~\bibnamefont {Lloyd}},\
  and\ \bibinfo {author} {\bibfnamefont {L.}~\bibnamefont {Maccone}},\
  }\href@noop {} {\bibfield  {journal} {\bibinfo  {journal} {Nat. Photonics}\
  }\textbf {\bibinfo {volume} {5}},\ \bibinfo {pages} {222} (\bibinfo {year}
  {2011})}\BibitemShut {NoStop}%
\bibitem [{\citenamefont {Demkowicz-Dobrzański}\ \emph
  {et~al.}(2012)\citenamefont {Demkowicz-Dobrzański}, \citenamefont
  {Kołodyński},\ and\ \citenamefont {Guţă}}]{Demkowicz2012-nat}%
  \BibitemOpen
  \bibfield  {author} {\bibinfo {author} {\bibfnamefont {R.}~\bibnamefont
  {Demkowicz-Dobrzański}}, \bibinfo {author} {\bibfnamefont {J.}~\bibnamefont
  {Kołodyński}},\ and\ \bibinfo {author} {\bibfnamefont {M.}~\bibnamefont
  {Guţă}},\ }\href@noop {} {\bibfield  {journal} {\bibinfo  {journal} {Nature
  Communications}\ }\textbf {\bibinfo {volume} {3}} (\bibinfo {year}
  {2012})}\BibitemShut {NoStop}%
\bibitem [{\citenamefont {Paris}(2009)}]{paris2009quantum}%
  \BibitemOpen
  \bibfield  {author} {\bibinfo {author} {\bibfnamefont {M.~G.}\ \bibnamefont
  {Paris}},\ }\href {https://doi.org/10.1142/S0219749909004839} {\bibfield
  {journal} {\bibinfo  {journal} {International Journal of Quantum
  Information}\ }\textbf {\bibinfo {volume} {7}},\ \bibinfo {pages} {125}
  (\bibinfo {year} {2009})}\BibitemShut {NoStop}%
\bibitem [{\citenamefont {Helstrom}(1969)}]{helstrom1969quantum}%
  \BibitemOpen
  \bibfield  {author} {\bibinfo {author} {\bibfnamefont {C.~W.}\ \bibnamefont
  {Helstrom}},\ }\href@noop {} {\bibfield  {journal} {\bibinfo  {journal} {J.
  Stat. Phys.}\ }\textbf {\bibinfo {volume} {1}},\ \bibinfo {pages} {231}
  (\bibinfo {year} {1969})}\BibitemShut {NoStop}%
\bibitem [{\citenamefont {Cram{\'e}r}(1999)}]{cramer1999mathematical}%
  \BibitemOpen
  \bibfield  {author} {\bibinfo {author} {\bibfnamefont {H.}~\bibnamefont
  {Cram{\'e}r}},\ }\href@noop {} {\emph {\bibinfo {title} {Mathematical methods
  of statistics}}},\ Vol.~\bibinfo {volume} {26}\ (\bibinfo  {publisher}
  {Princeton university press},\ \bibinfo {year} {1999})\BibitemShut {NoStop}%
\bibitem [{\citenamefont {Le~Cam}(1986)}]{LeCam-1986}%
  \BibitemOpen
  \bibfield  {author} {\bibinfo {author} {\bibfnamefont {L.~M.}\ \bibnamefont
  {Le~Cam}},\ }\href@noop {} {\emph {\bibinfo {title} {Asymptotic methods in
  statistical decision theory}}},\ Springer series in statistics\ (\bibinfo
  {publisher} {Springer-Verlag},\ \bibinfo {address} {New York},\ \bibinfo
  {year} {1986})\BibitemShut {NoStop}%
\bibitem [{\citenamefont {Lehmann}\ and\ \citenamefont
  {Casella}(2006)}]{lehmann2006theory}%
  \BibitemOpen
  \bibfield  {author} {\bibinfo {author} {\bibfnamefont {E.~L.}\ \bibnamefont
  {Lehmann}}\ and\ \bibinfo {author} {\bibfnamefont {G.}~\bibnamefont
  {Casella}},\ }\href@noop {} {\emph {\bibinfo {title} {Theory of point
  estimation}}}\ (\bibinfo  {publisher} {Springer Science \& Business Media},\
  \bibinfo {year} {2006})\BibitemShut {NoStop}%
\bibitem [{\citenamefont {Holevo}(1984)}]{Holevo}%
  \BibitemOpen
  \bibfield  {author} {\bibinfo {author} {\bibfnamefont {A.}~\bibnamefont
  {Holevo}},\ }in\ \href@noop {} {\emph {\bibinfo {booktitle} {Quantum
  Probability and Applications to the Quantum Theory of Irreversible
  Processes}}}\ (\bibinfo  {publisher} {Springer},\ \bibinfo {year} {1984})\
  pp.\ \bibinfo {pages} {153--172}\BibitemShut {NoStop}%
\bibitem [{\citenamefont {Nielsen}\ and\ \citenamefont
  {Chuang}(2000)}]{nielsen00}%
  \BibitemOpen
  \bibfield  {author} {\bibinfo {author} {\bibfnamefont {M.~A.}\ \bibnamefont
  {Nielsen}}\ and\ \bibinfo {author} {\bibfnamefont {I.~L.}\ \bibnamefont
  {Chuang}},\ }\href@noop {} {\emph {\bibinfo {title} {Quantum Computation and
  Quantum Information}}}\ (\bibinfo  {publisher} {Cambridge University Press},\
  \bibinfo {year} {2000})\BibitemShut {NoStop}%
\bibitem [{\citenamefont {Liu}\ \emph {et~al.}(2019)\citenamefont {Liu},
  \citenamefont {Yuan}, \citenamefont {Lu},\ and\ \citenamefont
  {Wang}}]{Liu_2019}%
  \BibitemOpen
  \bibfield  {author} {\bibinfo {author} {\bibfnamefont {J.}~\bibnamefont
  {Liu}}, \bibinfo {author} {\bibfnamefont {H.}~\bibnamefont {Yuan}}, \bibinfo
  {author} {\bibfnamefont {X.-M.}\ \bibnamefont {Lu}},\ and\ \bibinfo {author}
  {\bibfnamefont {X.}~\bibnamefont {Wang}},\ }\href
  {https://doi.org/10.1088/1751-8121/ab5d4d} {\bibfield  {journal} {\bibinfo
  {journal} {Journal of Physics A: Mathematical and Theoretical}\ }\textbf
  {\bibinfo {volume} {53}},\ \bibinfo {pages} {023001} (\bibinfo {year}
  {2019})}\BibitemShut {NoStop}%
\bibitem [{\citenamefont {Albarelli}\ \emph {et~al.}(2020)\citenamefont
  {Albarelli}, \citenamefont {Barbieri}, \citenamefont {Genoni},\ and\
  \citenamefont {Gianani}}]{Albarelli-2020}%
  \BibitemOpen
  \bibfield  {author} {\bibinfo {author} {\bibfnamefont {F.}~\bibnamefont
  {Albarelli}}, \bibinfo {author} {\bibfnamefont {M.}~\bibnamefont {Barbieri}},
  \bibinfo {author} {\bibfnamefont {M.}~\bibnamefont {Genoni}},\ and\ \bibinfo
  {author} {\bibfnamefont {I.}~\bibnamefont {Gianani}},\ }\href
  {https://doi.org/https://doi.org/10.1016/j.physleta.2020.126311} {\bibfield
  {journal} {\bibinfo  {journal} {Physics Letters A}\ }\textbf {\bibinfo
  {volume} {384}},\ \bibinfo {pages} {126311} (\bibinfo {year}
  {2020})}\BibitemShut {NoStop}%
\bibitem [{\citenamefont {Szczykulska}\ \emph {et~al.}(2016)\citenamefont
  {Szczykulska}, \citenamefont {Baumgratz},\ and\ \citenamefont
  {Datta}}]{szczykulska2016multi}%
  \BibitemOpen
  \bibfield  {author} {\bibinfo {author} {\bibfnamefont {M.}~\bibnamefont
  {Szczykulska}}, \bibinfo {author} {\bibfnamefont {T.}~\bibnamefont
  {Baumgratz}},\ and\ \bibinfo {author} {\bibfnamefont {A.}~\bibnamefont
  {Datta}},\ }\href@noop {} {\bibfield  {journal} {\bibinfo  {journal}
  {Advances in Physics: X}\ }\textbf {\bibinfo {volume} {1}},\ \bibinfo {pages}
  {621} (\bibinfo {year} {2016})}\BibitemShut {NoStop}%
\bibitem [{\citenamefont {Albarelli}\ \emph {et~al.}(2024)\citenamefont
  {Albarelli}, \citenamefont {Gianani}, \citenamefont {Genoni},\ and\
  \citenamefont {Barbieri}}]{albarelli2023fisher}%
  \BibitemOpen
  \bibfield  {author} {\bibinfo {author} {\bibfnamefont {F.}~\bibnamefont
  {Albarelli}}, \bibinfo {author} {\bibfnamefont {I.}~\bibnamefont {Gianani}},
  \bibinfo {author} {\bibfnamefont {M.~G.}\ \bibnamefont {Genoni}},\ and\
  \bibinfo {author} {\bibfnamefont {M.}~\bibnamefont {Barbieri}},\ }\href
  {https://doi.org/10.1103/PhysRevA.110.032436} {\bibfield  {journal} {\bibinfo
   {journal} {Phys. Rev. A}\ }\textbf {\bibinfo {volume} {110}},\ \bibinfo
  {pages} {032436} (\bibinfo {year} {2024})}\BibitemShut {NoStop}%
\bibitem [{\citenamefont {Baumgratz}\ and\ \citenamefont
  {Datta}(2016)}]{baumgratz2016quantum}%
  \BibitemOpen
  \bibfield  {author} {\bibinfo {author} {\bibfnamefont {T.}~\bibnamefont
  {Baumgratz}}\ and\ \bibinfo {author} {\bibfnamefont {A.}~\bibnamefont
  {Datta}},\ }\href {https://doi.org/10.1103/PhysRevLett.116.030801} {\bibfield
   {journal} {\bibinfo  {journal} {Phys. Rev. Lett.}\ }\textbf {\bibinfo
  {volume} {116}},\ \bibinfo {pages} {030801} (\bibinfo {year}
  {2016})}\BibitemShut {NoStop}%
\bibitem [{\citenamefont {Humphreys}\ \emph
  {et~al.}(2013{\natexlab{a}})\citenamefont {Humphreys}, \citenamefont
  {Barbieri}, \citenamefont {Datta},\ and\ \citenamefont
  {Walmsley}}]{humphreys2013quantum}%
  \BibitemOpen
  \bibfield  {author} {\bibinfo {author} {\bibfnamefont {P.~C.}\ \bibnamefont
  {Humphreys}}, \bibinfo {author} {\bibfnamefont {M.}~\bibnamefont {Barbieri}},
  \bibinfo {author} {\bibfnamefont {A.}~\bibnamefont {Datta}},\ and\ \bibinfo
  {author} {\bibfnamefont {I.~A.}\ \bibnamefont {Walmsley}},\ }\href
  {https://doi.org/10.1103/PhysRevLett.111.070403} {\bibfield  {journal}
  {\bibinfo  {journal} {Phys. Rev. Lett.}\ }\textbf {\bibinfo {volume} {111}},\
  \bibinfo {pages} {070403} (\bibinfo {year} {2013}{\natexlab{a}})}\BibitemShut
  {NoStop}%
\bibitem [{\citenamefont {Pezz\`e}\ \emph
  {et~al.}(2017{\natexlab{a}})\citenamefont {Pezz\`e}, \citenamefont
  {Ciampini}, \citenamefont {Spagnolo}, \citenamefont {Humphreys},
  \citenamefont {Datta}, \citenamefont {Walmsley}, \citenamefont {Barbieri},
  \citenamefont {Sciarrino},\ and\ \citenamefont {Smerzi}}]{pezze2017optimal}%
  \BibitemOpen
  \bibfield  {author} {\bibinfo {author} {\bibfnamefont {L.}~\bibnamefont
  {Pezz\`e}}, \bibinfo {author} {\bibfnamefont {M.~A.}\ \bibnamefont
  {Ciampini}}, \bibinfo {author} {\bibfnamefont {N.}~\bibnamefont {Spagnolo}},
  \bibinfo {author} {\bibfnamefont {P.~C.}\ \bibnamefont {Humphreys}}, \bibinfo
  {author} {\bibfnamefont {A.}~\bibnamefont {Datta}}, \bibinfo {author}
  {\bibfnamefont {I.~A.}\ \bibnamefont {Walmsley}}, \bibinfo {author}
  {\bibfnamefont {M.}~\bibnamefont {Barbieri}}, \bibinfo {author}
  {\bibfnamefont {F.}~\bibnamefont {Sciarrino}},\ and\ \bibinfo {author}
  {\bibfnamefont {A.}~\bibnamefont {Smerzi}},\ }\href
  {https://doi.org/10.1103/PhysRevLett.119.130504} {\bibfield  {journal}
  {\bibinfo  {journal} {Phys. Rev. Lett.}\ }\textbf {\bibinfo {volume} {119}},\
  \bibinfo {pages} {130504} (\bibinfo {year} {2017}{\natexlab{a}})}\BibitemShut
  {NoStop}%
\bibitem [{\citenamefont {Apellaniz}\ \emph {et~al.}(2018)\citenamefont
  {Apellaniz}, \citenamefont {Urizar-Lanz}, \citenamefont {Zimbor\'as},
  \citenamefont {Hyllus},\ and\ \citenamefont
  {T\'oth}}]{apellaniz2018precision}%
  \BibitemOpen
  \bibfield  {author} {\bibinfo {author} {\bibfnamefont {I.}~\bibnamefont
  {Apellaniz}}, \bibinfo {author} {\bibfnamefont {I.~n.}\ \bibnamefont
  {Urizar-Lanz}}, \bibinfo {author} {\bibfnamefont {Z.}~\bibnamefont
  {Zimbor\'as}}, \bibinfo {author} {\bibfnamefont {P.}~\bibnamefont {Hyllus}},\
  and\ \bibinfo {author} {\bibfnamefont {G.}~\bibnamefont {T\'oth}},\ }\href
  {https://doi.org/10.1103/PhysRevA.97.053603} {\bibfield  {journal} {\bibinfo
  {journal} {Phys. Rev. A}\ }\textbf {\bibinfo {volume} {97}},\ \bibinfo
  {pages} {053603} (\bibinfo {year} {2018})}\BibitemShut {NoStop}%
\bibitem [{\citenamefont {Genoni}\ \emph {et~al.}(2013)\citenamefont {Genoni},
  \citenamefont {Paris}, \citenamefont {Adesso}, \citenamefont {Nha},
  \citenamefont {Knight},\ and\ \citenamefont {Kim}}]{genoni2013optimal}%
  \BibitemOpen
  \bibfield  {author} {\bibinfo {author} {\bibfnamefont {M.~G.}\ \bibnamefont
  {Genoni}}, \bibinfo {author} {\bibfnamefont {M.~G.~A.}\ \bibnamefont
  {Paris}}, \bibinfo {author} {\bibfnamefont {G.}~\bibnamefont {Adesso}},
  \bibinfo {author} {\bibfnamefont {H.}~\bibnamefont {Nha}}, \bibinfo {author}
  {\bibfnamefont {P.~L.}\ \bibnamefont {Knight}},\ and\ \bibinfo {author}
  {\bibfnamefont {M.~S.}\ \bibnamefont {Kim}},\ }\href
  {https://doi.org/10.1103/PhysRevA.87.012107} {\bibfield  {journal} {\bibinfo
  {journal} {Phys. Rev. A}\ }\textbf {\bibinfo {volume} {87}},\ \bibinfo
  {pages} {012107} (\bibinfo {year} {2013})}\BibitemShut {NoStop}%
\bibitem [{\citenamefont {Bradshaw}\ \emph {et~al.}(2018)\citenamefont
  {Bradshaw}, \citenamefont {Lam},\ and\ \citenamefont
  {Assad}}]{PhysRevA.97.012106}%
  \BibitemOpen
  \bibfield  {author} {\bibinfo {author} {\bibfnamefont {M.}~\bibnamefont
  {Bradshaw}}, \bibinfo {author} {\bibfnamefont {P.~K.}\ \bibnamefont {Lam}},\
  and\ \bibinfo {author} {\bibfnamefont {S.~M.}\ \bibnamefont {Assad}},\ }\href
  {https://doi.org/10.1103/PhysRevA.97.012106} {\bibfield  {journal} {\bibinfo
  {journal} {Phys. Rev. A}\ }\textbf {\bibinfo {volume} {97}},\ \bibinfo
  {pages} {012106} (\bibinfo {year} {2018})}\BibitemShut {NoStop}%
\bibitem [{\citenamefont {Humphreys}\ \emph
  {et~al.}(2013{\natexlab{b}})\citenamefont {Humphreys}, \citenamefont
  {Barbieri}, \citenamefont {Datta},\ and\ \citenamefont
  {Walmsley}}]{PhysRevLett.111.070403}%
  \BibitemOpen
  \bibfield  {author} {\bibinfo {author} {\bibfnamefont {P.~C.}\ \bibnamefont
  {Humphreys}}, \bibinfo {author} {\bibfnamefont {M.}~\bibnamefont {Barbieri}},
  \bibinfo {author} {\bibfnamefont {A.}~\bibnamefont {Datta}},\ and\ \bibinfo
  {author} {\bibfnamefont {I.~A.}\ \bibnamefont {Walmsley}},\ }\href
  {https://doi.org/10.1103/PhysRevLett.111.070403} {\bibfield  {journal}
  {\bibinfo  {journal} {Phys. Rev. Lett.}\ }\textbf {\bibinfo {volume} {111}},\
  \bibinfo {pages} {070403} (\bibinfo {year} {2013}{\natexlab{b}})}\BibitemShut
  {NoStop}%
\bibitem [{\citenamefont {Gagatsos}\ \emph {et~al.}(2016)\citenamefont
  {Gagatsos}, \citenamefont {Branford},\ and\ \citenamefont
  {Datta}}]{PhysRevA.94.042342}%
  \BibitemOpen
  \bibfield  {author} {\bibinfo {author} {\bibfnamefont {C.~N.}\ \bibnamefont
  {Gagatsos}}, \bibinfo {author} {\bibfnamefont {D.}~\bibnamefont {Branford}},\
  and\ \bibinfo {author} {\bibfnamefont {A.}~\bibnamefont {Datta}},\ }\href
  {https://doi.org/10.1103/PhysRevA.94.042342} {\bibfield  {journal} {\bibinfo
  {journal} {Phys. Rev. A}\ }\textbf {\bibinfo {volume} {94}},\ \bibinfo
  {pages} {042342} (\bibinfo {year} {2016})}\BibitemShut {NoStop}%
\bibitem [{\citenamefont {Knott}\ \emph {et~al.}(2016)\citenamefont {Knott},
  \citenamefont {Proctor}, \citenamefont {Hayes}, \citenamefont {Ralph},
  \citenamefont {Kok},\ and\ \citenamefont {Dunningham}}]{PhysRevA.94.062312}%
  \BibitemOpen
  \bibfield  {author} {\bibinfo {author} {\bibfnamefont {P.~A.}\ \bibnamefont
  {Knott}}, \bibinfo {author} {\bibfnamefont {T.~J.}\ \bibnamefont {Proctor}},
  \bibinfo {author} {\bibfnamefont {A.~J.}\ \bibnamefont {Hayes}}, \bibinfo
  {author} {\bibfnamefont {J.~F.}\ \bibnamefont {Ralph}}, \bibinfo {author}
  {\bibfnamefont {P.}~\bibnamefont {Kok}},\ and\ \bibinfo {author}
  {\bibfnamefont {J.~A.}\ \bibnamefont {Dunningham}},\ }\href
  {https://doi.org/10.1103/PhysRevA.94.062312} {\bibfield  {journal} {\bibinfo
  {journal} {Phys. Rev. A}\ }\textbf {\bibinfo {volume} {94}},\ \bibinfo
  {pages} {062312} (\bibinfo {year} {2016})}\BibitemShut {NoStop}%
\bibitem [{\citenamefont {Pezz\`e}\ \emph
  {et~al.}(2017{\natexlab{b}})\citenamefont {Pezz\`e}, \citenamefont
  {Ciampini}, \citenamefont {Spagnolo}, \citenamefont {Humphreys},
  \citenamefont {Datta}, \citenamefont {Walmsley}, \citenamefont {Barbieri},
  \citenamefont {Sciarrino},\ and\ \citenamefont
  {Smerzi}}]{PhysRevLett.119.130504}%
  \BibitemOpen
  \bibfield  {author} {\bibinfo {author} {\bibfnamefont {L.}~\bibnamefont
  {Pezz\`e}}, \bibinfo {author} {\bibfnamefont {M.~A.}\ \bibnamefont
  {Ciampini}}, \bibinfo {author} {\bibfnamefont {N.}~\bibnamefont {Spagnolo}},
  \bibinfo {author} {\bibfnamefont {P.~C.}\ \bibnamefont {Humphreys}}, \bibinfo
  {author} {\bibfnamefont {A.}~\bibnamefont {Datta}}, \bibinfo {author}
  {\bibfnamefont {I.~A.}\ \bibnamefont {Walmsley}}, \bibinfo {author}
  {\bibfnamefont {M.}~\bibnamefont {Barbieri}}, \bibinfo {author}
  {\bibfnamefont {F.}~\bibnamefont {Sciarrino}},\ and\ \bibinfo {author}
  {\bibfnamefont {A.}~\bibnamefont {Smerzi}},\ }\href
  {https://doi.org/10.1103/PhysRevLett.119.130504} {\bibfield  {journal}
  {\bibinfo  {journal} {Phys. Rev. Lett.}\ }\textbf {\bibinfo {volume} {119}},\
  \bibinfo {pages} {130504} (\bibinfo {year} {2017}{\natexlab{b}})}\BibitemShut
  {NoStop}%
\bibitem [{\citenamefont {Vidrighin}\ \emph {et~al.}(2014)\citenamefont
  {Vidrighin}, \citenamefont {Donati}, \citenamefont {Genoni}, \citenamefont
  {Jin}, \citenamefont {Kolthammer}, \citenamefont {Kim}, \citenamefont
  {Datta}, \citenamefont {Barbieri},\ and\ \citenamefont
  {Walmsley}}]{vidrighin2014joint}%
  \BibitemOpen
  \bibfield  {author} {\bibinfo {author} {\bibfnamefont {M.}~\bibnamefont
  {Vidrighin}}, \bibinfo {author} {\bibfnamefont {G.}~\bibnamefont {Donati}},
  \bibinfo {author} {\bibfnamefont {M.}~\bibnamefont {Genoni}}, \bibinfo
  {author} {\bibfnamefont {X.}~\bibnamefont {Jin}}, \bibinfo {author}
  {\bibfnamefont {W.}~\bibnamefont {Kolthammer}}, \bibinfo {author}
  {\bibfnamefont {M.}~\bibnamefont {Kim}}, \bibinfo {author} {\bibfnamefont
  {A.}~\bibnamefont {Datta}}, \bibinfo {author} {\bibfnamefont
  {M.}~\bibnamefont {Barbieri}},\ and\ \bibinfo {author} {\bibfnamefont
  {I.}~\bibnamefont {Walmsley}},\ }\href@noop {} {\bibfield  {journal}
  {\bibinfo  {journal} {Commun}\ }\textbf {\bibinfo {volume} {5}},\ \bibinfo
  {pages} {3532} (\bibinfo {year} {2014})}\BibitemShut {NoStop}%
\bibitem [{\citenamefont {Lu}\ and\ \citenamefont
  {Wang}(2021)}]{PhysRevLett.126.120503}%
  \BibitemOpen
  \bibfield  {author} {\bibinfo {author} {\bibfnamefont {X.-M.}\ \bibnamefont
  {Lu}}\ and\ \bibinfo {author} {\bibfnamefont {X.}~\bibnamefont {Wang}},\
  }\href {https://doi.org/10.1103/PhysRevLett.126.120503} {\bibfield  {journal}
  {\bibinfo  {journal} {Phys. Rev. Lett.}\ }\textbf {\bibinfo {volume} {126}},\
  \bibinfo {pages} {120503} (\bibinfo {year} {2021})}\BibitemShut {NoStop}%
\bibitem [{\citenamefont {Carollo}\ \emph {et~al.}(2019)\citenamefont
  {Carollo}, \citenamefont {Spagnolo}, \citenamefont {Dubkov},\ and\
  \citenamefont {Valenti}}]{Carollo_2019}%
  \BibitemOpen
  \bibfield  {author} {\bibinfo {author} {\bibfnamefont {A.}~\bibnamefont
  {Carollo}}, \bibinfo {author} {\bibfnamefont {B.}~\bibnamefont {Spagnolo}},
  \bibinfo {author} {\bibfnamefont {A.~A.}\ \bibnamefont {Dubkov}},\ and\
  \bibinfo {author} {\bibfnamefont {D.}~\bibnamefont {Valenti}},\ }\href
  {https://doi.org/10.1088/1742-5468/ab3ccb} {\bibfield  {journal} {\bibinfo
  {journal} {Journal of Statistical Mechanics: Theory and Experiment}\ }\textbf
  {\bibinfo {volume} {2019}},\ \bibinfo {pages} {094010} (\bibinfo {year}
  {2019})}\BibitemShut {NoStop}%
\bibitem [{\citenamefont {Di~Fresco}\ \emph {et~al.}(2022)\citenamefont
  {Di~Fresco}, \citenamefont {Spagnolo}, \citenamefont {Valenti},\ and\
  \citenamefont {Carollo}}]{di2022multiparameter}%
  \BibitemOpen
  \bibfield  {author} {\bibinfo {author} {\bibfnamefont {G.}~\bibnamefont
  {Di~Fresco}}, \bibinfo {author} {\bibfnamefont {B.}~\bibnamefont {Spagnolo}},
  \bibinfo {author} {\bibfnamefont {D.}~\bibnamefont {Valenti}},\ and\ \bibinfo
  {author} {\bibfnamefont {A.}~\bibnamefont {Carollo}},\ }\href@noop {}
  {\bibfield  {journal} {\bibinfo  {journal} {SciPost Physics}\ }\textbf
  {\bibinfo {volume} {13}},\ \bibinfo {pages} {077} (\bibinfo {year}
  {2022})}\BibitemShut {NoStop}%
\bibitem [{\citenamefont {Ragy}\ \emph {et~al.}(2016)\citenamefont {Ragy},
  \citenamefont {Jarzyna},\ and\ \citenamefont
  {Demkowicz-Dobrza\ifmmode~\acute{n}\else
  \'{n}\fi{}ski}}]{ragy2016compatibility}%
  \BibitemOpen
  \bibfield  {author} {\bibinfo {author} {\bibfnamefont {S.}~\bibnamefont
  {Ragy}}, \bibinfo {author} {\bibfnamefont {M.}~\bibnamefont {Jarzyna}},\ and\
  \bibinfo {author} {\bibfnamefont {R.}~\bibnamefont
  {Demkowicz-Dobrza\ifmmode~\acute{n}\else \'{n}\fi{}ski}},\ }\href
  {https://doi.org/10.1103/PhysRevA.94.052108} {\bibfield  {journal} {\bibinfo
  {journal} {Phys. Rev. A}\ }\textbf {\bibinfo {volume} {94}},\ \bibinfo
  {pages} {052108} (\bibinfo {year} {2016})}\BibitemShut {NoStop}%
\bibitem [{\citenamefont {Demkowicz-Dobrzański}\ \emph
  {et~al.}(2020)\citenamefont {Demkowicz-Dobrzański}, \citenamefont
  {Górecki},\ and\ \citenamefont {Guţă}}]{Demkowicz-Dobrzański_2020}%
  \BibitemOpen
  \bibfield  {author} {\bibinfo {author} {\bibfnamefont {R.}~\bibnamefont
  {Demkowicz-Dobrzański}}, \bibinfo {author} {\bibfnamefont {W.}~\bibnamefont
  {Górecki}},\ and\ \bibinfo {author} {\bibfnamefont {M.}~\bibnamefont
  {Guţă}},\ }\href {https://doi.org/10.1088/1751-8121/ab8ef3} {\bibfield
  {journal} {\bibinfo  {journal} {Journal of Physics A: Mathematical and
  Theoretical}\ }\textbf {\bibinfo {volume} {53}},\ \bibinfo {pages} {363001}
  (\bibinfo {year} {2020})}\BibitemShut {NoStop}%
\bibitem [{\citenamefont {Belliardo}\ and\ \citenamefont
  {Giovannetti}(2021)}]{Belliardo_2021}%
  \BibitemOpen
  \bibfield  {author} {\bibinfo {author} {\bibfnamefont {F.}~\bibnamefont
  {Belliardo}}\ and\ \bibinfo {author} {\bibfnamefont {V.}~\bibnamefont
  {Giovannetti}},\ }\href {https://doi.org/10.1088/1367-2630/ac04ca} {\bibfield
   {journal} {\bibinfo  {journal} {New Journal of Physics}\ }\textbf {\bibinfo
  {volume} {23}},\ \bibinfo {pages} {063055} (\bibinfo {year}
  {2021})}\BibitemShut {NoStop}%
\bibitem [{\citenamefont {Namkung}\ \emph {et~al.}(2024)\citenamefont
  {Namkung}, \citenamefont {Lee},\ and\ \citenamefont
  {Lim}}]{namkung2024unifiedcramerraoboundquantum}%
  \BibitemOpen
  \bibfield  {author} {\bibinfo {author} {\bibfnamefont {M.}~\bibnamefont
  {Namkung}}, \bibinfo {author} {\bibfnamefont {C.}~\bibnamefont {Lee}},\ and\
  \bibinfo {author} {\bibfnamefont {H.-T.}\ \bibnamefont {Lim}},\ }\href
  {https://arxiv.org/abs/2412.01117} {\bibinfo {title} {Unified cram\'{e}r-rao
  bound for quantum multi-parameter estimation: Invertible and non-invertible
  fisher information matrix}} (\bibinfo {year} {2024}),\ \Eprint
  {https://arxiv.org/abs/2412.01117} {arXiv:2412.01117 [quant-ph]} \BibitemShut
  {NoStop}%
\bibitem [{\citenamefont {Mihailescu}\ \emph
  {et~al.}(2024{\natexlab{a}})\citenamefont {Mihailescu}, \citenamefont
  {Campbell},\ and\ \citenamefont {Gietka}}]{mihailescu2024uncertain}%
  \BibitemOpen
  \bibfield  {author} {\bibinfo {author} {\bibfnamefont {G.}~\bibnamefont
  {Mihailescu}}, \bibinfo {author} {\bibfnamefont {S.}~\bibnamefont
  {Campbell}},\ and\ \bibinfo {author} {\bibfnamefont {K.}~\bibnamefont
  {Gietka}},\ }\href@noop {} {\bibfield  {journal} {\bibinfo  {journal} {arXiv
  preprint arXiv:2407.19917}\ } (\bibinfo {year}
  {2024}{\natexlab{a}})}\BibitemShut {NoStop}%
\bibitem [{\citenamefont {Goldberg}\ \emph {et~al.}(2021)\citenamefont
  {Goldberg}, \citenamefont {Romero}, \citenamefont {Sanz},\ and\ \citenamefont
  {S{\'a}nchez-Soto}}]{goldberg2021taming}%
  \BibitemOpen
  \bibfield  {author} {\bibinfo {author} {\bibfnamefont {A.~Z.}\ \bibnamefont
  {Goldberg}}, \bibinfo {author} {\bibfnamefont {J.~L.}\ \bibnamefont
  {Romero}}, \bibinfo {author} {\bibfnamefont {{\'A}.~S.}\ \bibnamefont
  {Sanz}},\ and\ \bibinfo {author} {\bibfnamefont {L.~L.}\ \bibnamefont
  {S{\'a}nchez-Soto}},\ }\href@noop {} {\bibfield  {journal} {\bibinfo
  {journal} {International Journal of Quantum Information}\ }\textbf {\bibinfo
  {volume} {19}},\ \bibinfo {pages} {2140004} (\bibinfo {year}
  {2021})}\BibitemShut {NoStop}%
\bibitem [{\citenamefont {Mihailescu}\ \emph
  {et~al.}(2024{\natexlab{b}})\citenamefont {Mihailescu}, \citenamefont
  {Bayat}, \citenamefont {Campbell},\ and\ \citenamefont
  {Mitchell}}]{Mihailescu_2024}%
  \BibitemOpen
  \bibfield  {author} {\bibinfo {author} {\bibfnamefont {G.}~\bibnamefont
  {Mihailescu}}, \bibinfo {author} {\bibfnamefont {A.}~\bibnamefont {Bayat}},
  \bibinfo {author} {\bibfnamefont {S.}~\bibnamefont {Campbell}},\ and\
  \bibinfo {author} {\bibfnamefont {A.~K.}\ \bibnamefont {Mitchell}},\ }\href
  {https://doi.org/10.1088/2058-9565/ad438d} {\bibfield  {journal} {\bibinfo
  {journal} {Quantum Science and Technology}\ }\textbf {\bibinfo {volume}
  {9}},\ \bibinfo {pages} {035033} (\bibinfo {year}
  {2024}{\natexlab{b}})}\BibitemShut {NoStop}%
\bibitem [{\citenamefont {Candeloro}\ \emph {et~al.}(2024)\citenamefont
  {Candeloro}, \citenamefont {Pazhotan},\ and\ \citenamefont
  {Paris}}]{Candeloro_2024}%
  \BibitemOpen
  \bibfield  {author} {\bibinfo {author} {\bibfnamefont {A.}~\bibnamefont
  {Candeloro}}, \bibinfo {author} {\bibfnamefont {Z.}~\bibnamefont
  {Pazhotan}},\ and\ \bibinfo {author} {\bibfnamefont {M.~G.~A.}\ \bibnamefont
  {Paris}},\ }\href {https://doi.org/10.1088/2058-9565/ad7498} {\bibfield
  {journal} {\bibinfo  {journal} {Quantum Science and Technology}\ }\textbf
  {\bibinfo {volume} {9}},\ \bibinfo {pages} {045045} (\bibinfo {year}
  {2024})}\BibitemShut {NoStop}%
\bibitem [{\citenamefont {Burgarth}\ \emph {et~al.}(2015)\citenamefont
  {Burgarth}, \citenamefont {Giovannetti}, \citenamefont {Kato},\ and\
  \citenamefont {Yuasa}}]{burgarth2015quantum}%
  \BibitemOpen
  \bibfield  {author} {\bibinfo {author} {\bibfnamefont {D.}~\bibnamefont
  {Burgarth}}, \bibinfo {author} {\bibfnamefont {V.}~\bibnamefont
  {Giovannetti}}, \bibinfo {author} {\bibfnamefont {A.~N.}\ \bibnamefont
  {Kato}},\ and\ \bibinfo {author} {\bibfnamefont {K.}~\bibnamefont {Yuasa}},\
  }\href@noop {} {\bibfield  {journal} {\bibinfo  {journal} {New Journal of
  Physics}\ }\textbf {\bibinfo {volume} {17}},\ \bibinfo {pages} {113055}
  (\bibinfo {year} {2015})}\BibitemShut {NoStop}%
\bibitem [{\citenamefont {Montenegro}\ \emph {et~al.}(2022)\citenamefont
  {Montenegro}, \citenamefont {Jones}, \citenamefont {Bose},\ and\
  \citenamefont {Bayat}}]{montenegro2022sequential}%
  \BibitemOpen
  \bibfield  {author} {\bibinfo {author} {\bibfnamefont {V.}~\bibnamefont
  {Montenegro}}, \bibinfo {author} {\bibfnamefont {G.~S.}\ \bibnamefont
  {Jones}}, \bibinfo {author} {\bibfnamefont {S.}~\bibnamefont {Bose}},\ and\
  \bibinfo {author} {\bibfnamefont {A.}~\bibnamefont {Bayat}},\ }\href
  {https://doi.org/10.1103/PhysRevLett.129.120503} {\bibfield  {journal}
  {\bibinfo  {journal} {Phys. Rev. Lett.}\ }\textbf {\bibinfo {volume} {129}},\
  \bibinfo {pages} {120503} (\bibinfo {year} {2022})}\BibitemShut {NoStop}%
\bibitem [{\citenamefont {Yang}\ \emph {et~al.}(2023)\citenamefont {Yang},
  \citenamefont {Montenegro},\ and\ \citenamefont
  {Bayat}}]{PhysRevResearch.5.043273}%
  \BibitemOpen
  \bibfield  {author} {\bibinfo {author} {\bibfnamefont {Y.}~\bibnamefont
  {Yang}}, \bibinfo {author} {\bibfnamefont {V.}~\bibnamefont {Montenegro}},\
  and\ \bibinfo {author} {\bibfnamefont {A.}~\bibnamefont {Bayat}},\ }\href
  {https://doi.org/10.1103/PhysRevResearch.5.043273} {\bibfield  {journal}
  {\bibinfo  {journal} {Phys. Rev. Res.}\ }\textbf {\bibinfo {volume} {5}},\
  \bibinfo {pages} {043273} (\bibinfo {year} {2023})}\BibitemShut {NoStop}%
\bibitem [{\citenamefont {Bompais}\ \emph {et~al.}(2022)\citenamefont
  {Bompais}, \citenamefont {Amini},\ and\ \citenamefont
  {Pellegrini}}]{bompais2022parameter}%
  \BibitemOpen
  \bibfield  {author} {\bibinfo {author} {\bibfnamefont {M.}~\bibnamefont
  {Bompais}}, \bibinfo {author} {\bibfnamefont {N.~H.}\ \bibnamefont {Amini}},\
  and\ \bibinfo {author} {\bibfnamefont {C.}~\bibnamefont {Pellegrini}},\ }in\
  \href@noop {} {\emph {\bibinfo {booktitle} {2022 IEEE 61st Conference on
  Decision and Control (CDC)}}}\ (\bibinfo {organization} {IEEE},\ \bibinfo
  {year} {2022})\ pp.\ \bibinfo {pages} {5161--5166}\BibitemShut {NoStop}%
\bibitem [{\citenamefont {Clark}\ \emph {et~al.}(2019)\citenamefont {Clark},
  \citenamefont {Stokes},\ and\ \citenamefont {Beige}}]{clark2019quantum}%
  \BibitemOpen
  \bibfield  {author} {\bibinfo {author} {\bibfnamefont {L.~A.}\ \bibnamefont
  {Clark}}, \bibinfo {author} {\bibfnamefont {A.}~\bibnamefont {Stokes}},\ and\
  \bibinfo {author} {\bibfnamefont {A.}~\bibnamefont {Beige}},\ }\href@noop {}
  {\bibfield  {journal} {\bibinfo  {journal} {Physical Review A}\ }\textbf
  {\bibinfo {volume} {99}},\ \bibinfo {pages} {022102} (\bibinfo {year}
  {2019})}\BibitemShut {NoStop}%
\bibitem [{\citenamefont {Ma}\ \emph {et~al.}(2018)\citenamefont {Ma},
  \citenamefont {Wang}, \citenamefont {Leong},\ and\ \citenamefont
  {Liu}}]{ma2018phase}%
  \BibitemOpen
  \bibfield  {author} {\bibinfo {author} {\bibfnamefont {W.-L.}\ \bibnamefont
  {Ma}}, \bibinfo {author} {\bibfnamefont {P.}~\bibnamefont {Wang}}, \bibinfo
  {author} {\bibfnamefont {W.-H.}\ \bibnamefont {Leong}},\ and\ \bibinfo
  {author} {\bibfnamefont {R.-B.}\ \bibnamefont {Liu}},\ }\href@noop {}
  {\bibfield  {journal} {\bibinfo  {journal} {Physical Review A}\ }\textbf
  {\bibinfo {volume} {98}},\ \bibinfo {pages} {012117} (\bibinfo {year}
  {2018})}\BibitemShut {NoStop}%
\bibitem [{\citenamefont {Benoist}\ \emph {et~al.}(2019)\citenamefont
  {Benoist}, \citenamefont {Fraas}, \citenamefont {Pautrat},\ and\
  \citenamefont {Pellegrini}}]{benoist2019invariant}%
  \BibitemOpen
  \bibfield  {author} {\bibinfo {author} {\bibfnamefont {T.}~\bibnamefont
  {Benoist}}, \bibinfo {author} {\bibfnamefont {M.}~\bibnamefont {Fraas}},
  \bibinfo {author} {\bibfnamefont {Y.}~\bibnamefont {Pautrat}},\ and\ \bibinfo
  {author} {\bibfnamefont {C.}~\bibnamefont {Pellegrini}},\ }\href@noop {}
  {\bibfield  {journal} {\bibinfo  {journal} {Probability Theory and Related
  Fields}\ }\textbf {\bibinfo {volume} {174}},\ \bibinfo {pages} {307}
  (\bibinfo {year} {2019})}\BibitemShut {NoStop}%
\bibitem [{\citenamefont {Benoist}\ \emph {et~al.}(2023)\citenamefont
  {Benoist}, \citenamefont {Fatras},\ and\ \citenamefont
  {Pellegrini}}]{benoist2023limit}%
  \BibitemOpen
  \bibfield  {author} {\bibinfo {author} {\bibfnamefont {T.}~\bibnamefont
  {Benoist}}, \bibinfo {author} {\bibfnamefont {J.-L.}\ \bibnamefont
  {Fatras}},\ and\ \bibinfo {author} {\bibfnamefont {C.}~\bibnamefont
  {Pellegrini}},\ }\href@noop {} {\bibfield  {journal} {\bibinfo  {journal}
  {Stochastic Processes and their Applications}\ }\textbf {\bibinfo {volume}
  {164}},\ \bibinfo {pages} {288} (\bibinfo {year} {2023})}\BibitemShut
  {NoStop}%
\bibitem [{\citenamefont {Block}\ \emph {et~al.}(2022)\citenamefont {Block},
  \citenamefont {Bao}, \citenamefont {Choi}, \citenamefont {Altman},\ and\
  \citenamefont {Yao}}]{PhysRevLett.128.010604}%
  \BibitemOpen
  \bibfield  {author} {\bibinfo {author} {\bibfnamefont {M.}~\bibnamefont
  {Block}}, \bibinfo {author} {\bibfnamefont {Y.}~\bibnamefont {Bao}}, \bibinfo
  {author} {\bibfnamefont {S.}~\bibnamefont {Choi}}, \bibinfo {author}
  {\bibfnamefont {E.}~\bibnamefont {Altman}},\ and\ \bibinfo {author}
  {\bibfnamefont {N.~Y.}\ \bibnamefont {Yao}},\ }\href@noop {} {\bibfield
  {journal} {\bibinfo  {journal} {Phys. Rev. Lett.}\ }\textbf {\bibinfo
  {volume} {128}},\ \bibinfo {pages} {010604} (\bibinfo {year}
  {2022})}\BibitemShut {NoStop}%
\bibitem [{\citenamefont {Skinner}\ \emph {et~al.}(2019)\citenamefont
  {Skinner}, \citenamefont {Ruhman},\ and\ \citenamefont
  {Nahum}}]{PhysRevX.9.031009}%
  \BibitemOpen
  \bibfield  {author} {\bibinfo {author} {\bibfnamefont {B.}~\bibnamefont
  {Skinner}}, \bibinfo {author} {\bibfnamefont {J.}~\bibnamefont {Ruhman}},\
  and\ \bibinfo {author} {\bibfnamefont {A.}~\bibnamefont {Nahum}},\ }\href
  {https://doi.org/10.1103/PhysRevX.9.031009} {\bibfield  {journal} {\bibinfo
  {journal} {Phys. Rev. X}\ }\textbf {\bibinfo {volume} {9}},\ \bibinfo {pages}
  {031009} (\bibinfo {year} {2019})}\BibitemShut {NoStop}%
\bibitem [{\citenamefont {Meyer}(2021)}]{Meyer_2021}%
  \BibitemOpen
  \bibfield  {author} {\bibinfo {author} {\bibfnamefont {J.~J.}\ \bibnamefont
  {Meyer}},\ }\href {https://doi.org/10.22331/q-2021-09-09-539} {\bibfield
  {journal} {\bibinfo  {journal} {Quantum}\ }\textbf {\bibinfo {volume} {5}},\
  \bibinfo {pages} {539} (\bibinfo {year} {2021})}\BibitemShut {NoStop}%
\bibitem [{\citenamefont {Stoica}\ and\ \citenamefont
  {Marzetta}(2001)}]{stoica2001parameter}%
  \BibitemOpen
  \bibfield  {author} {\bibinfo {author} {\bibfnamefont {P.}~\bibnamefont
  {Stoica}}\ and\ \bibinfo {author} {\bibfnamefont {T.}~\bibnamefont
  {Marzetta}},\ }\href {https://doi.org/10.1109/78.890346} {\bibfield
  {journal} {\bibinfo  {journal} {IEEE Transactions on Signal Processing}\
  }\textbf {\bibinfo {volume} {49}},\ \bibinfo {pages} {87} (\bibinfo {year}
  {2001})}\BibitemShut {NoStop}%
\bibitem [{SM_()}]{SM_multiparameter}%
  \BibitemOpen
  \href@noop {} {}\bibinfo {note} {Supplemental Material includes}\BibitemShut
  {NoStop}%
\bibitem [{\citenamefont {Holevo}(2011)}]{holevo2011probabilistic}%
  \BibitemOpen
  \bibfield  {author} {\bibinfo {author} {\bibfnamefont {A.~S.}\ \bibnamefont
  {Holevo}},\ }\href@noop {} {\emph {\bibinfo {title} {Probabilistic and
  statistical aspects of quantum theory}}},\ Vol.~\bibinfo {volume} {1}\
  (\bibinfo  {publisher} {Springer Science \& Business Media},\ \bibinfo {year}
  {2011})\BibitemShut {NoStop}%
\bibitem [{\citenamefont {Albarelli}\ \emph {et~al.}(2019)\citenamefont
  {Albarelli}, \citenamefont {Friel},\ and\ \citenamefont
  {Datta}}]{albarelli2019evaluating}%
  \BibitemOpen
  \bibfield  {author} {\bibinfo {author} {\bibfnamefont {F.}~\bibnamefont
  {Albarelli}}, \bibinfo {author} {\bibfnamefont {J.~F.}\ \bibnamefont
  {Friel}},\ and\ \bibinfo {author} {\bibfnamefont {A.}~\bibnamefont {Datta}},\
  }\href@noop {} {\bibfield  {journal} {\bibinfo  {journal} {Physical review
  letters}\ }\textbf {\bibinfo {volume} {123}},\ \bibinfo {pages} {200503}
  (\bibinfo {year} {2019})}\BibitemShut {NoStop}%
\bibitem [{\citenamefont {Munkres}(2018)}]{munkres2018elements}%
  \BibitemOpen
  \bibfield  {author} {\bibinfo {author} {\bibfnamefont {J.~R.}\ \bibnamefont
  {Munkres}},\ }\href@noop {} {\emph {\bibinfo {title} {Elements of algebraic
  topology}}}\ (\bibinfo  {publisher} {CRC press},\ \bibinfo {year}
  {2018})\BibitemShut {NoStop}%
\bibitem [{\citenamefont {Yang}\ \emph {et~al.}(2024)\citenamefont {Yang},
  \citenamefont {Montenegro},\ and\ \citenamefont
  {Bayat}}]{yang2024sequential}%
  \BibitemOpen
  \bibfield  {author} {\bibinfo {author} {\bibfnamefont {Y.}~\bibnamefont
  {Yang}}, \bibinfo {author} {\bibfnamefont {V.}~\bibnamefont {Montenegro}},\
  and\ \bibinfo {author} {\bibfnamefont {A.}~\bibnamefont {Bayat}},\ }\href
  {https://doi.org/10.1103/PhysRevApplied.22.024069} {\bibfield  {journal}
  {\bibinfo  {journal} {Phys. Rev. Appl.}\ }\textbf {\bibinfo {volume} {22}},\
  \bibinfo {pages} {024069} (\bibinfo {year} {2024})}\BibitemShut {NoStop}%
\bibitem [{\citenamefont {Lieb}\ and\ \citenamefont
  {Robinson}(1972)}]{lieb1972finite}%
  \BibitemOpen
  \bibfield  {author} {\bibinfo {author} {\bibfnamefont {E.~H.}\ \bibnamefont
  {Lieb}}\ and\ \bibinfo {author} {\bibfnamefont {D.~W.}\ \bibnamefont
  {Robinson}},\ }\href {https://doi.org/10.1007/BF01645779} {\bibfield
  {journal} {\bibinfo  {journal} {Communications in Mathematical Physics}\
  }\textbf {\bibinfo {volume} {28}},\ \bibinfo {pages} {251} (\bibinfo {year}
  {1972})}\BibitemShut {NoStop}%
\bibitem [{\citenamefont {Facchi}\ and\ \citenamefont
  {Pascazio}(2008)}]{facchi2008quantum}%
  \BibitemOpen
  \bibfield  {author} {\bibinfo {author} {\bibfnamefont {P.}~\bibnamefont
  {Facchi}}\ and\ \bibinfo {author} {\bibfnamefont {S.}~\bibnamefont
  {Pascazio}},\ }\href {https://doi.org/10.1088/1751-8113/41/49/493001}
  {\bibfield  {journal} {\bibinfo  {journal} {Journal of Physics A:
  Mathematical and Theoretical}\ }\textbf {\bibinfo {volume} {41}},\ \bibinfo
  {pages} {493001} (\bibinfo {year} {2008})}\BibitemShut {NoStop}%
\bibitem [{\citenamefont {Ding}\ \emph {et~al.}(2022)\citenamefont {Ding},
  \citenamefont {Liu}, \citenamefont {Shi}, \citenamefont {Guo}, \citenamefont
  {M{\o}lmer},\ and\ \citenamefont {Adams}}]{Ding2022}%
  \BibitemOpen
  \bibfield  {author} {\bibinfo {author} {\bibfnamefont {D.-S.}\ \bibnamefont
  {Ding}}, \bibinfo {author} {\bibfnamefont {Z.-K.}\ \bibnamefont {Liu}},
  \bibinfo {author} {\bibfnamefont {B.-S.}\ \bibnamefont {Shi}}, \bibinfo
  {author} {\bibfnamefont {G.-C.}\ \bibnamefont {Guo}}, \bibinfo {author}
  {\bibfnamefont {K.}~\bibnamefont {M{\o}lmer}},\ and\ \bibinfo {author}
  {\bibfnamefont {C.~S.}\ \bibnamefont {Adams}},\ }\href@noop {} {\bibfield
  {journal} {\bibinfo  {journal} {Nature Physics}\ }\textbf {\bibinfo {volume}
  {18}},\ \bibinfo {pages} {1447} (\bibinfo {year} {2022})}\BibitemShut
  {NoStop}%
\bibitem [{\citenamefont {Liu}\ \emph {et~al.}(2021)\citenamefont {Liu},
  \citenamefont {Chen}, \citenamefont {Jiang}, \citenamefont {Yang},
  \citenamefont {Wu}, \citenamefont {Li}, \citenamefont {Yuan}, \citenamefont
  {Peng},\ and\ \citenamefont {Du}}]{Liu2021}%
  \BibitemOpen
  \bibfield  {author} {\bibinfo {author} {\bibfnamefont {R.}~\bibnamefont
  {Liu}}, \bibinfo {author} {\bibfnamefont {Y.}~\bibnamefont {Chen}}, \bibinfo
  {author} {\bibfnamefont {M.}~\bibnamefont {Jiang}}, \bibinfo {author}
  {\bibfnamefont {X.}~\bibnamefont {Yang}}, \bibinfo {author} {\bibfnamefont
  {Z.}~\bibnamefont {Wu}}, \bibinfo {author} {\bibfnamefont {Y.}~\bibnamefont
  {Li}}, \bibinfo {author} {\bibfnamefont {H.}~\bibnamefont {Yuan}}, \bibinfo
  {author} {\bibfnamefont {X.}~\bibnamefont {Peng}},\ and\ \bibinfo {author}
  {\bibfnamefont {J.}~\bibnamefont {Du}},\ }\href
  {https://doi.org/10.1038/s41534-021-00507-x} {\bibfield  {journal} {\bibinfo
  {journal} {npj Quantum Information}\ }\textbf {\bibinfo {volume} {7}},\
  \bibinfo {pages} {170} (\bibinfo {year} {2021})}\BibitemShut {NoStop}%
\bibitem [{\citenamefont {Zanardi}\ \emph {et~al.}(2008)\citenamefont
  {Zanardi}, \citenamefont {Paris},\ and\ \citenamefont
  {Venuti}}]{zanardi2008quantum}%
  \BibitemOpen
  \bibfield  {author} {\bibinfo {author} {\bibfnamefont {P.}~\bibnamefont
  {Zanardi}}, \bibinfo {author} {\bibfnamefont {M.~G.}\ \bibnamefont {Paris}},\
  and\ \bibinfo {author} {\bibfnamefont {L.~C.}\ \bibnamefont {Venuti}},\
  }\href@noop {} {\bibfield  {journal} {\bibinfo  {journal} {Phys. Rev. A}\
  }\textbf {\bibinfo {volume} {78}},\ \bibinfo {pages} {042105} (\bibinfo
  {year} {2008})}\BibitemShut {NoStop}%
\bibitem [{\citenamefont {Invernizzi}\ \emph {et~al.}(2008)\citenamefont
  {Invernizzi}, \citenamefont {Korbman}, \citenamefont {Venuti},\ and\
  \citenamefont {Paris}}]{invernizzi2008optimal}%
  \BibitemOpen
  \bibfield  {author} {\bibinfo {author} {\bibfnamefont {C.}~\bibnamefont
  {Invernizzi}}, \bibinfo {author} {\bibfnamefont {M.}~\bibnamefont {Korbman}},
  \bibinfo {author} {\bibfnamefont {L.~C.}\ \bibnamefont {Venuti}},\ and\
  \bibinfo {author} {\bibfnamefont {M.~G.}\ \bibnamefont {Paris}},\ }\href@noop
  {} {\bibfield  {journal} {\bibinfo  {journal} {Phys. Rev. A}\ }\textbf
  {\bibinfo {volume} {78}},\ \bibinfo {pages} {042106} (\bibinfo {year}
  {2008})}\BibitemShut {NoStop}%
\bibitem [{\citenamefont {Salvatori}\ \emph {et~al.}(2014)\citenamefont
  {Salvatori}, \citenamefont {Mandarino},\ and\ \citenamefont
  {Paris}}]{salvatori2014quantum}%
  \BibitemOpen
  \bibfield  {author} {\bibinfo {author} {\bibfnamefont {G.}~\bibnamefont
  {Salvatori}}, \bibinfo {author} {\bibfnamefont {A.}~\bibnamefont
  {Mandarino}},\ and\ \bibinfo {author} {\bibfnamefont {M.~G.}\ \bibnamefont
  {Paris}},\ }\href@noop {} {\bibfield  {journal} {\bibinfo  {journal} {Phys.
  Rev. A}\ }\textbf {\bibinfo {volume} {90}},\ \bibinfo {pages} {022111}
  (\bibinfo {year} {2014})}\BibitemShut {NoStop}%
\bibitem [{\citenamefont {Zanardi}\ \emph {et~al.}(2007)\citenamefont
  {Zanardi}, \citenamefont {Quan}, \citenamefont {Wang},\ and\ \citenamefont
  {Sun}}]{zanardi2007critical}%
  \BibitemOpen
  \bibfield  {author} {\bibinfo {author} {\bibfnamefont {P.}~\bibnamefont
  {Zanardi}}, \bibinfo {author} {\bibfnamefont {H.~T.}\ \bibnamefont {Quan}},
  \bibinfo {author} {\bibfnamefont {X.}~\bibnamefont {Wang}},\ and\ \bibinfo
  {author} {\bibfnamefont {C.~P.}\ \bibnamefont {Sun}},\ }\href@noop {}
  {\bibfield  {journal} {\bibinfo  {journal} {Phys. Rev. A}\ }\textbf {\bibinfo
  {volume} {75}},\ \bibinfo {pages} {032109} (\bibinfo {year}
  {2007})}\BibitemShut {NoStop}%
\bibitem [{\citenamefont {Garbe}\ \emph
  {et~al.}(2020{\natexlab{a}})\citenamefont {Garbe}, \citenamefont {Bina},
  \citenamefont {Keller}, \citenamefont {Paris},\ and\ \citenamefont
  {Felicetti}}]{Garbe2020}%
  \BibitemOpen
  \bibfield  {author} {\bibinfo {author} {\bibfnamefont {L.}~\bibnamefont
  {Garbe}}, \bibinfo {author} {\bibfnamefont {M.}~\bibnamefont {Bina}},
  \bibinfo {author} {\bibfnamefont {A.}~\bibnamefont {Keller}}, \bibinfo
  {author} {\bibfnamefont {M.~G.~A.}\ \bibnamefont {Paris}},\ and\ \bibinfo
  {author} {\bibfnamefont {S.}~\bibnamefont {Felicetti}},\ }\href
  {https://doi.org/10.1103/PhysRevLett.124.120504} {\bibfield  {journal}
  {\bibinfo  {journal} {Phys. Rev. Lett.}\ }\textbf {\bibinfo {volume} {124}},\
  \bibinfo {pages} {120504} (\bibinfo {year} {2020}{\natexlab{a}})}\BibitemShut
  {NoStop}%
\bibitem [{\citenamefont {Chu}\ \emph {et~al.}(2021)\citenamefont {Chu},
  \citenamefont {Zhang}, \citenamefont {Yu},\ and\ \citenamefont
  {Cai}}]{Jianming2021critical}%
  \BibitemOpen
  \bibfield  {author} {\bibinfo {author} {\bibfnamefont {Y.}~\bibnamefont
  {Chu}}, \bibinfo {author} {\bibfnamefont {S.}~\bibnamefont {Zhang}}, \bibinfo
  {author} {\bibfnamefont {B.}~\bibnamefont {Yu}},\ and\ \bibinfo {author}
  {\bibfnamefont {J.}~\bibnamefont {Cai}},\ }\href@noop {} {\bibfield
  {journal} {\bibinfo  {journal} {Phys. Rev. Lett.}\ }\textbf {\bibinfo
  {volume} {126}},\ \bibinfo {pages} {010502} (\bibinfo {year}
  {2021})}\BibitemShut {NoStop}%
\bibitem [{\citenamefont {Rams}\ \emph
  {et~al.}(2018{\natexlab{a}})\citenamefont {Rams}, \citenamefont {Sierant},
  \citenamefont {Dutta}, \citenamefont {Horodecki},\ and\ \citenamefont
  {Zakrzewski}}]{Horodecki2018prx}%
  \BibitemOpen
  \bibfield  {author} {\bibinfo {author} {\bibfnamefont {M.~M.}\ \bibnamefont
  {Rams}}, \bibinfo {author} {\bibfnamefont {P.}~\bibnamefont {Sierant}},
  \bibinfo {author} {\bibfnamefont {O.}~\bibnamefont {Dutta}}, \bibinfo
  {author} {\bibfnamefont {P.}~\bibnamefont {Horodecki}},\ and\ \bibinfo
  {author} {\bibfnamefont {J.}~\bibnamefont {Zakrzewski}},\ }\href@noop {}
  {\bibfield  {journal} {\bibinfo  {journal} {Phys. Rev. X}\ }\textbf {\bibinfo
  {volume} {8}},\ \bibinfo {pages} {021022} (\bibinfo {year}
  {2018}{\natexlab{a}})}\BibitemShut {NoStop}%
\bibitem [{\citenamefont {Sarkar}\ \emph {et~al.}(2022)\citenamefont {Sarkar},
  \citenamefont {Mukhopadhyay}, \citenamefont {Alase},\ and\ \citenamefont
  {Bayat}}]{Sarkar2022topological}%
  \BibitemOpen
  \bibfield  {author} {\bibinfo {author} {\bibfnamefont {S.}~\bibnamefont
  {Sarkar}}, \bibinfo {author} {\bibfnamefont {C.}~\bibnamefont
  {Mukhopadhyay}}, \bibinfo {author} {\bibfnamefont {A.}~\bibnamefont
  {Alase}},\ and\ \bibinfo {author} {\bibfnamefont {A.}~\bibnamefont {Bayat}},\
  }\href {https://doi.org/10.1103/PhysRevLett.129.090503} {\bibfield  {journal}
  {\bibinfo  {journal} {Phys. Rev. Lett.}\ }\textbf {\bibinfo {volume} {129}},\
  \bibinfo {pages} {090503} (\bibinfo {year} {2022})}\BibitemShut {NoStop}%
\bibitem [{\citenamefont {Rams}\ \emph
  {et~al.}(2018{\natexlab{b}})\citenamefont {Rams}, \citenamefont {Sierant},
  \citenamefont {Dutta}, \citenamefont {Horodecki},\ and\ \citenamefont
  {Zakrzewski}}]{PhysRevX.8.021022}%
  \BibitemOpen
  \bibfield  {author} {\bibinfo {author} {\bibfnamefont {M.~M.}\ \bibnamefont
  {Rams}}, \bibinfo {author} {\bibfnamefont {P.}~\bibnamefont {Sierant}},
  \bibinfo {author} {\bibfnamefont {O.}~\bibnamefont {Dutta}}, \bibinfo
  {author} {\bibfnamefont {P.}~\bibnamefont {Horodecki}},\ and\ \bibinfo
  {author} {\bibfnamefont {J.}~\bibnamefont {Zakrzewski}},\ }\href@noop {}
  {\bibfield  {journal} {\bibinfo  {journal} {Phys. Rev. X}\ }\textbf {\bibinfo
  {volume} {8}},\ \bibinfo {pages} {021022} (\bibinfo {year}
  {2018}{\natexlab{b}})}\BibitemShut {NoStop}%
\bibitem [{\citenamefont {Montenegro}\ \emph {et~al.}(2021)\citenamefont
  {Montenegro}, \citenamefont {Mishra},\ and\ \citenamefont
  {Bayat}}]{Montenegro2021}%
  \BibitemOpen
  \bibfield  {author} {\bibinfo {author} {\bibfnamefont {V.}~\bibnamefont
  {Montenegro}}, \bibinfo {author} {\bibfnamefont {U.}~\bibnamefont {Mishra}},\
  and\ \bibinfo {author} {\bibfnamefont {A.}~\bibnamefont {Bayat}},\ }\href
  {https://doi.org/10.1103/PhysRevLett.126.200501} {\bibfield  {journal}
  {\bibinfo  {journal} {Phys. Rev. Lett.}\ }\textbf {\bibinfo {volume} {126}},\
  \bibinfo {pages} {200501} (\bibinfo {year} {2021})}\BibitemShut {NoStop}%
\bibitem [{\citenamefont {Fern{\'a}ndez-Lorenzo}\ and\ \citenamefont
  {Porras}(2017)}]{fernandez2017quantum}%
  \BibitemOpen
  \bibfield  {author} {\bibinfo {author} {\bibfnamefont {S.}~\bibnamefont
  {Fern{\'a}ndez-Lorenzo}}\ and\ \bibinfo {author} {\bibfnamefont
  {D.}~\bibnamefont {Porras}},\ }\href
  {https://link.aps.org/doi/10.1103/PhysRevA.96.013817} {\bibfield  {journal}
  {\bibinfo  {journal} {Phys. Rev. A}\ }\textbf {\bibinfo {volume} {96}},\
  \bibinfo {pages} {013817} (\bibinfo {year} {2017})}\BibitemShut {NoStop}%
\bibitem [{\citenamefont {Baumann}\ \emph {et~al.}(2010)\citenamefont
  {Baumann}, \citenamefont {Guerlin}, \citenamefont {Brennecke},\ and\
  \citenamefont {Esslinger}}]{baumann2010dicke}%
  \BibitemOpen
  \bibfield  {author} {\bibinfo {author} {\bibfnamefont {K.}~\bibnamefont
  {Baumann}}, \bibinfo {author} {\bibfnamefont {C.}~\bibnamefont {Guerlin}},
  \bibinfo {author} {\bibfnamefont {F.}~\bibnamefont {Brennecke}},\ and\
  \bibinfo {author} {\bibfnamefont {T.}~\bibnamefont {Esslinger}},\ }\href
  {https://www.nature.com/articles/nature09009} {\bibfield  {journal} {\bibinfo
   {journal} {Nature}\ }\textbf {\bibinfo {volume} {464}},\ \bibinfo {pages}
  {1301} (\bibinfo {year} {2010})}\BibitemShut {NoStop}%
\bibitem [{\citenamefont {Baden}\ \emph {et~al.}(2014)\citenamefont {Baden},
  \citenamefont {Arnold}, \citenamefont {Grimsmo}, \citenamefont {Parkins},\
  and\ \citenamefont {Barrett}}]{baden2014realization}%
  \BibitemOpen
  \bibfield  {author} {\bibinfo {author} {\bibfnamefont {M.~P.}\ \bibnamefont
  {Baden}}, \bibinfo {author} {\bibfnamefont {K.~J.}\ \bibnamefont {Arnold}},
  \bibinfo {author} {\bibfnamefont {A.~L.}\ \bibnamefont {Grimsmo}}, \bibinfo
  {author} {\bibfnamefont {S.}~\bibnamefont {Parkins}},\ and\ \bibinfo {author}
  {\bibfnamefont {M.~D.}\ \bibnamefont {Barrett}},\ }\href
  {https://link.aps.org/doi/10.1103/PhysRevLett.113.020408} {\bibfield
  {journal} {\bibinfo  {journal} {Phys. Rev. Lett.}\ }\textbf {\bibinfo
  {volume} {113}},\ \bibinfo {pages} {020408} (\bibinfo {year}
  {2014})}\BibitemShut {NoStop}%
\bibitem [{\citenamefont {Klinder}\ \emph {et~al.}(2015)\citenamefont
  {Klinder}, \citenamefont {Ke{\ss}ler}, \citenamefont {Wolke}, \citenamefont
  {Mathey},\ and\ \citenamefont {Hemmerich}}]{klinder2015dynamical}%
  \BibitemOpen
  \bibfield  {author} {\bibinfo {author} {\bibfnamefont {J.}~\bibnamefont
  {Klinder}}, \bibinfo {author} {\bibfnamefont {H.}~\bibnamefont {Ke{\ss}ler}},
  \bibinfo {author} {\bibfnamefont {M.}~\bibnamefont {Wolke}}, \bibinfo
  {author} {\bibfnamefont {L.}~\bibnamefont {Mathey}},\ and\ \bibinfo {author}
  {\bibfnamefont {A.}~\bibnamefont {Hemmerich}},\ }\href
  {https://doi.org/10.1073/pnas.1417132112} {\bibfield  {journal} {\bibinfo
  {journal} {Proc. Natl. Acad. Sci. U.S.A.}\ }\textbf {\bibinfo {volume}
  {112}},\ \bibinfo {pages} {3290} (\bibinfo {year} {2015})}\BibitemShut
  {NoStop}%
\bibitem [{\citenamefont {Rodriguez}\ \emph {et~al.}(2017)\citenamefont
  {Rodriguez}, \citenamefont {Casteels}, \citenamefont {Storme}, \citenamefont
  {Zambon}, \citenamefont {Sagnes}, \citenamefont {Le~Gratiet}, \citenamefont
  {Galopin}, \citenamefont {Lema{\^\i}tre}, \citenamefont {Amo}, \citenamefont
  {Ciuti} \emph {et~al.}}]{rodriguez2017probing}%
  \BibitemOpen
  \bibfield  {author} {\bibinfo {author} {\bibfnamefont {S.}~\bibnamefont
  {Rodriguez}}, \bibinfo {author} {\bibfnamefont {W.}~\bibnamefont {Casteels}},
  \bibinfo {author} {\bibfnamefont {F.}~\bibnamefont {Storme}}, \bibinfo
  {author} {\bibfnamefont {N.~C.}\ \bibnamefont {Zambon}}, \bibinfo {author}
  {\bibfnamefont {I.}~\bibnamefont {Sagnes}}, \bibinfo {author} {\bibfnamefont
  {L.}~\bibnamefont {Le~Gratiet}}, \bibinfo {author} {\bibfnamefont
  {E.}~\bibnamefont {Galopin}}, \bibinfo {author} {\bibfnamefont
  {A.}~\bibnamefont {Lema{\^\i}tre}}, \bibinfo {author} {\bibfnamefont
  {A.}~\bibnamefont {Amo}}, \bibinfo {author} {\bibfnamefont {C.}~\bibnamefont
  {Ciuti}}, \emph {et~al.},\ }\href
  {https://link.aps.org/doi/10.1103/PhysRevLett.118.247402} {\bibfield
  {journal} {\bibinfo  {journal} {Phys. Rev. Lett.}\ }\textbf {\bibinfo
  {volume} {118}},\ \bibinfo {pages} {247402} (\bibinfo {year}
  {2017})}\BibitemShut {NoStop}%
\bibitem [{\citenamefont {Fitzpatrick}\ \emph {et~al.}(2017)\citenamefont
  {Fitzpatrick}, \citenamefont {Sundaresan}, \citenamefont {Li}, \citenamefont
  {Koch},\ and\ \citenamefont {Houck}}]{fitzpatrick2017observation}%
  \BibitemOpen
  \bibfield  {author} {\bibinfo {author} {\bibfnamefont {M.}~\bibnamefont
  {Fitzpatrick}}, \bibinfo {author} {\bibfnamefont {N.~M.}\ \bibnamefont
  {Sundaresan}}, \bibinfo {author} {\bibfnamefont {A.~C.}\ \bibnamefont {Li}},
  \bibinfo {author} {\bibfnamefont {J.}~\bibnamefont {Koch}},\ and\ \bibinfo
  {author} {\bibfnamefont {A.~A.}\ \bibnamefont {Houck}},\ }\href
  {https://link.aps.org/doi/10.1103/PhysRevX.7.011016} {\bibfield  {journal}
  {\bibinfo  {journal} {Phys. Rev. X}\ }\textbf {\bibinfo {volume} {7}},\
  \bibinfo {pages} {011016} (\bibinfo {year} {2017})}\BibitemShut {NoStop}%
\bibitem [{\citenamefont {Fink}\ \emph {et~al.}(2017)\citenamefont {Fink},
  \citenamefont {Dombi}, \citenamefont {Vukics}, \citenamefont {Wallraff},\
  and\ \citenamefont {Domokos}}]{fink2017observation}%
  \BibitemOpen
  \bibfield  {author} {\bibinfo {author} {\bibfnamefont {J.~M.}\ \bibnamefont
  {Fink}}, \bibinfo {author} {\bibfnamefont {A.}~\bibnamefont {Dombi}},
  \bibinfo {author} {\bibfnamefont {A.}~\bibnamefont {Vukics}}, \bibinfo
  {author} {\bibfnamefont {A.}~\bibnamefont {Wallraff}},\ and\ \bibinfo
  {author} {\bibfnamefont {P.}~\bibnamefont {Domokos}},\ }\href
  {https://link.aps.org/doi/10.1103/PhysRevX.7.011012} {\bibfield  {journal}
  {\bibinfo  {journal} {Phys. Rev. X}\ }\textbf {\bibinfo {volume} {7}},\
  \bibinfo {pages} {011012} (\bibinfo {year} {2017})}\BibitemShut {NoStop}%
\bibitem [{\citenamefont {Ilias}\ \emph {et~al.}(2022)\citenamefont {Ilias},
  \citenamefont {Yang}, \citenamefont {Huelga},\ and\ \citenamefont
  {Plenio}}]{ilias2022criticality}%
  \BibitemOpen
  \bibfield  {author} {\bibinfo {author} {\bibfnamefont {T.}~\bibnamefont
  {Ilias}}, \bibinfo {author} {\bibfnamefont {D.}~\bibnamefont {Yang}},
  \bibinfo {author} {\bibfnamefont {S.~F.}\ \bibnamefont {Huelga}},\ and\
  \bibinfo {author} {\bibfnamefont {M.~B.}\ \bibnamefont {Plenio}},\ }\href
  {https://link.aps.org/doi/10.1103/PRXQuantum.3.010354} {\bibfield  {journal}
  {\bibinfo  {journal} {PRX Quantum}\ }\textbf {\bibinfo {volume} {3}},\
  \bibinfo {pages} {010354} (\bibinfo {year} {2022})}\BibitemShut {NoStop}%
\bibitem [{\citenamefont {Ilias}\ \emph {et~al.}(2024)\citenamefont {Ilias},
  \citenamefont {Yang}, \citenamefont {Huelga},\ and\ \citenamefont
  {Plenio}}]{Ilias2023Criticality}%
  \BibitemOpen
  \bibfield  {author} {\bibinfo {author} {\bibfnamefont {T.}~\bibnamefont
  {Ilias}}, \bibinfo {author} {\bibfnamefont {D.}~\bibnamefont {Yang}},
  \bibinfo {author} {\bibfnamefont {S.~F.}\ \bibnamefont {Huelga}},\ and\
  \bibinfo {author} {\bibfnamefont {M.~B.}\ \bibnamefont {Plenio}},\ }\href
  {https://doi.org/10.1038/s41534-024-00833-w} {\bibfield  {journal} {\bibinfo
  {journal} {npj Quantum Inf.}\ }\textbf {\bibinfo {volume} {10}},\ \bibinfo
  {pages} {36} (\bibinfo {year} {2024})}\BibitemShut {NoStop}%
\bibitem [{\citenamefont {Alipour}\ \emph {et~al.}(2014)\citenamefont
  {Alipour}, \citenamefont {Mehboudi},\ and\ \citenamefont
  {Rezakhani}}]{Alipour2014Quantum}%
  \BibitemOpen
  \bibfield  {author} {\bibinfo {author} {\bibfnamefont {S.}~\bibnamefont
  {Alipour}}, \bibinfo {author} {\bibfnamefont {M.}~\bibnamefont {Mehboudi}},\
  and\ \bibinfo {author} {\bibfnamefont {A.~T.}\ \bibnamefont {Rezakhani}},\
  }\href {https://doi.org/10.1103/PhysRevLett.112.120405} {\bibfield  {journal}
  {\bibinfo  {journal} {Phys. Rev. Lett.}\ }\textbf {\bibinfo {volume} {112}},\
  \bibinfo {pages} {120405} (\bibinfo {year} {2014})}\BibitemShut {NoStop}%
\bibitem [{\citenamefont {Mishra}\ and\ \citenamefont
  {Bayat}(2021)}]{mishra2021driving}%
  \BibitemOpen
  \bibfield  {author} {\bibinfo {author} {\bibfnamefont {U.}~\bibnamefont
  {Mishra}}\ and\ \bibinfo {author} {\bibfnamefont {A.}~\bibnamefont {Bayat}},\
  }\href {https://link.aps.org/doi/10.1103/PhysRevLett.127.080504} {\bibfield
  {journal} {\bibinfo  {journal} {Phys. Rev. Lett.}\ }\textbf {\bibinfo
  {volume} {127}},\ \bibinfo {pages} {080504} (\bibinfo {year}
  {2021})}\BibitemShut {NoStop}%
\bibitem [{\citenamefont {Mishra}\ and\ \citenamefont
  {Bayat}(2022)}]{mishra2022integrable}%
  \BibitemOpen
  \bibfield  {author} {\bibinfo {author} {\bibfnamefont {U.}~\bibnamefont
  {Mishra}}\ and\ \bibinfo {author} {\bibfnamefont {A.}~\bibnamefont {Bayat}},\
  }\href {https://www.nature.com/articles/s41598-022-17381-y} {\bibfield
  {journal} {\bibinfo  {journal} {Sci. Rep.}\ }\textbf {\bibinfo {volume}
  {12}},\ \bibinfo {pages} {1} (\bibinfo {year} {2022})}\BibitemShut {NoStop}%
\bibitem [{\citenamefont {Montenegro}\ \emph {et~al.}(2023)\citenamefont
  {Montenegro}, \citenamefont {Genoni}, \citenamefont {Bayat},\ and\
  \citenamefont {Paris}}]{montenegro2023quantum}%
  \BibitemOpen
  \bibfield  {author} {\bibinfo {author} {\bibfnamefont {V.}~\bibnamefont
  {Montenegro}}, \bibinfo {author} {\bibfnamefont {M.~G.}\ \bibnamefont
  {Genoni}}, \bibinfo {author} {\bibfnamefont {A.}~\bibnamefont {Bayat}},\ and\
  \bibinfo {author} {\bibfnamefont {M.~G.}\ \bibnamefont {Paris}},\ }\href
  {https://www.nature.com/articles/s42005-023-01423-6} {\bibfield  {journal}
  {\bibinfo  {journal} {Commun. Phys.}\ }\textbf {\bibinfo {volume} {6}},\
  \bibinfo {pages} {304} (\bibinfo {year} {2023})}\BibitemShut {NoStop}%
\bibitem [{\citenamefont {Cabot}\ \emph {et~al.}(2024)\citenamefont {Cabot},
  \citenamefont {Carollo},\ and\ \citenamefont
  {Lesanovsky}}]{Cabot2024Continuous}%
  \BibitemOpen
  \bibfield  {author} {\bibinfo {author} {\bibfnamefont {A.}~\bibnamefont
  {Cabot}}, \bibinfo {author} {\bibfnamefont {F.}~\bibnamefont {Carollo}},\
  and\ \bibinfo {author} {\bibfnamefont {I.}~\bibnamefont {Lesanovsky}},\
  }\href {https://doi.org/10.1103/PhysRevLett.132.050801} {\bibfield  {journal}
  {\bibinfo  {journal} {Phys. Rev. Lett.}\ }\textbf {\bibinfo {volume} {132}},\
  \bibinfo {pages} {050801} (\bibinfo {year} {2024})}\BibitemShut {NoStop}%
\bibitem [{\citenamefont {Lyu}\ \emph {et~al.}(2020)\citenamefont {Lyu},
  \citenamefont {Choudhury}, \citenamefont {Lv}, \citenamefont {Yan},\ and\
  \citenamefont {Zhou}}]{lyu2020eternal}%
  \BibitemOpen
  \bibfield  {author} {\bibinfo {author} {\bibfnamefont {C.}~\bibnamefont
  {Lyu}}, \bibinfo {author} {\bibfnamefont {S.}~\bibnamefont {Choudhury}},
  \bibinfo {author} {\bibfnamefont {C.}~\bibnamefont {Lv}}, \bibinfo {author}
  {\bibfnamefont {Y.}~\bibnamefont {Yan}},\ and\ \bibinfo {author}
  {\bibfnamefont {Q.}~\bibnamefont {Zhou}},\ }\href
  {https://link.aps.org/doi/10.1103/PhysRevResearch.2.033070} {\bibfield
  {journal} {\bibinfo  {journal} {Phys. Rev. Res.}\ }\textbf {\bibinfo {volume}
  {2}},\ \bibinfo {pages} {033070} (\bibinfo {year} {2020})}\BibitemShut
  {NoStop}%
\bibitem [{\citenamefont {Iemini}\ \emph {et~al.}(2024)\citenamefont {Iemini},
  \citenamefont {Fazio},\ and\ \citenamefont {Sanpera}}]{Iemini2023}%
  \BibitemOpen
  \bibfield  {author} {\bibinfo {author} {\bibfnamefont {F.}~\bibnamefont
  {Iemini}}, \bibinfo {author} {\bibfnamefont {R.}~\bibnamefont {Fazio}},\ and\
  \bibinfo {author} {\bibfnamefont {A.}~\bibnamefont {Sanpera}},\ }\href
  {https://link.aps.org/doi/10.1103/PhysRevA.109.L050203} {\bibfield  {journal}
  {\bibinfo  {journal} {Phys. Rev. A}\ }\textbf {\bibinfo {volume} {109}},\
  \bibinfo {pages} {L050203} (\bibinfo {year} {2024})}\BibitemShut {NoStop}%
\bibitem [{\citenamefont {Yousefjani}\ \emph {et~al.}(2024)\citenamefont
  {Yousefjani}, \citenamefont {Sacha},\ and\ \citenamefont
  {Bayat}}]{yousefjani2024discrete}%
  \BibitemOpen
  \bibfield  {author} {\bibinfo {author} {\bibfnamefont {R.}~\bibnamefont
  {Yousefjani}}, \bibinfo {author} {\bibfnamefont {K.}~\bibnamefont {Sacha}},\
  and\ \bibinfo {author} {\bibfnamefont {A.}~\bibnamefont {Bayat}},\ }\href
  {https://doi.org/10.48550/arXiv.2405.00328} {\bibfield  {journal} {\bibinfo
  {journal} {arXiv:2405.00328}\ } (\bibinfo {year} {2024})}\BibitemShut
  {NoStop}%
\bibitem [{\citenamefont {Kucsko}\ \emph {et~al.}(2013)\citenamefont {Kucsko},
  \citenamefont {Maurer}, \citenamefont {Yao}, \citenamefont {Kubo},
  \citenamefont {Noh}, \citenamefont {Lo}, \citenamefont {Park},\ and\
  \citenamefont {Lukin}}]{kucsko2013nanometre}%
  \BibitemOpen
  \bibfield  {author} {\bibinfo {author} {\bibfnamefont {G.}~\bibnamefont
  {Kucsko}}, \bibinfo {author} {\bibfnamefont {P.~C.}\ \bibnamefont {Maurer}},
  \bibinfo {author} {\bibfnamefont {N.~Y.}\ \bibnamefont {Yao}}, \bibinfo
  {author} {\bibfnamefont {M.}~\bibnamefont {Kubo}}, \bibinfo {author}
  {\bibfnamefont {H.~J.}\ \bibnamefont {Noh}}, \bibinfo {author} {\bibfnamefont
  {P.~K.}\ \bibnamefont {Lo}}, \bibinfo {author} {\bibfnamefont
  {H.}~\bibnamefont {Park}},\ and\ \bibinfo {author} {\bibfnamefont {M.~D.}\
  \bibnamefont {Lukin}},\ }\href@noop {} {\bibfield  {journal} {\bibinfo
  {journal} {Nature}\ }\textbf {\bibinfo {volume} {500}},\ \bibinfo {pages}
  {54} (\bibinfo {year} {2013})}\BibitemShut {NoStop}%
\bibitem [{\citenamefont {Taylor}\ \emph {et~al.}(2008)\citenamefont {Taylor},
  \citenamefont {Cappellaro}, \citenamefont {Childress}, \citenamefont {Jiang},
  \citenamefont {Budker}, \citenamefont {Hemmer}, \citenamefont {Yacoby},
  \citenamefont {Walsworth},\ and\ \citenamefont {Lukin}}]{taylor2008high}%
  \BibitemOpen
  \bibfield  {author} {\bibinfo {author} {\bibfnamefont {J.~M.}\ \bibnamefont
  {Taylor}}, \bibinfo {author} {\bibfnamefont {P.}~\bibnamefont {Cappellaro}},
  \bibinfo {author} {\bibfnamefont {L.}~\bibnamefont {Childress}}, \bibinfo
  {author} {\bibfnamefont {L.}~\bibnamefont {Jiang}}, \bibinfo {author}
  {\bibfnamefont {D.}~\bibnamefont {Budker}}, \bibinfo {author} {\bibfnamefont
  {P.}~\bibnamefont {Hemmer}}, \bibinfo {author} {\bibfnamefont
  {A.}~\bibnamefont {Yacoby}}, \bibinfo {author} {\bibfnamefont
  {R.}~\bibnamefont {Walsworth}},\ and\ \bibinfo {author} {\bibfnamefont
  {M.}~\bibnamefont {Lukin}},\ }\href@noop {} {\bibfield  {journal} {\bibinfo
  {journal} {Nature Physics}\ }\textbf {\bibinfo {volume} {4}},\ \bibinfo
  {pages} {810} (\bibinfo {year} {2008})}\BibitemShut {NoStop}%
\bibitem [{\citenamefont {Garbe}\ \emph
  {et~al.}(2020{\natexlab{b}})\citenamefont {Garbe}, \citenamefont {Bina},
  \citenamefont {Keller}, \citenamefont {Paris},\ and\ \citenamefont
  {Felicetti}}]{garbe2020critical}%
  \BibitemOpen
  \bibfield  {author} {\bibinfo {author} {\bibfnamefont {L.}~\bibnamefont
  {Garbe}}, \bibinfo {author} {\bibfnamefont {M.}~\bibnamefont {Bina}},
  \bibinfo {author} {\bibfnamefont {A.}~\bibnamefont {Keller}}, \bibinfo
  {author} {\bibfnamefont {M.~G.}\ \bibnamefont {Paris}},\ and\ \bibinfo
  {author} {\bibfnamefont {S.}~\bibnamefont {Felicetti}},\ }\href@noop {}
  {\bibfield  {journal} {\bibinfo  {journal} {Physical review letters}\
  }\textbf {\bibinfo {volume} {124}},\ \bibinfo {pages} {120504} (\bibinfo
  {year} {2020}{\natexlab{b}})}\BibitemShut {NoStop}%
\bibitem [{\citenamefont {Barrett}\ \emph {et~al.}(2016)\citenamefont
  {Barrett}, \citenamefont {Bertoldi},\ and\ \citenamefont
  {Bouyer}}]{Barrett_2016}%
  \BibitemOpen
  \bibfield  {author} {\bibinfo {author} {\bibfnamefont {B.}~\bibnamefont
  {Barrett}}, \bibinfo {author} {\bibfnamefont {A.}~\bibnamefont {Bertoldi}},\
  and\ \bibinfo {author} {\bibfnamefont {P.}~\bibnamefont {Bouyer}},\ }\href
  {https://doi.org/10.1088/0031-8949/91/5/053006} {\bibfield  {journal}
  {\bibinfo  {journal} {Physica Scripta}\ }\textbf {\bibinfo {volume} {91}},\
  \bibinfo {pages} {053006} (\bibinfo {year} {2016})}\BibitemShut {NoStop}%
\bibitem [{\citenamefont {Karnieli}\ \emph {et~al.}(2023)\citenamefont
  {Karnieli}, \citenamefont {Tsesses}, \citenamefont {Yu}, \citenamefont
  {Rivera}, \citenamefont {Zhao}, \citenamefont {Arie}, \citenamefont {Fan},\
  and\ \citenamefont {Kaminer}}]{karnieli2023quantum}%
  \BibitemOpen
  \bibfield  {author} {\bibinfo {author} {\bibfnamefont {A.}~\bibnamefont
  {Karnieli}}, \bibinfo {author} {\bibfnamefont {S.}~\bibnamefont {Tsesses}},
  \bibinfo {author} {\bibfnamefont {R.}~\bibnamefont {Yu}}, \bibinfo {author}
  {\bibfnamefont {N.}~\bibnamefont {Rivera}}, \bibinfo {author} {\bibfnamefont
  {Z.}~\bibnamefont {Zhao}}, \bibinfo {author} {\bibfnamefont {A.}~\bibnamefont
  {Arie}}, \bibinfo {author} {\bibfnamefont {S.}~\bibnamefont {Fan}},\ and\
  \bibinfo {author} {\bibfnamefont {I.}~\bibnamefont {Kaminer}},\ }\href@noop
  {} {\bibfield  {journal} {\bibinfo  {journal} {Science advances}\ }\textbf
  {\bibinfo {volume} {9}},\ \bibinfo {pages} {eadd2349} (\bibinfo {year}
  {2023})}\BibitemShut {NoStop}%
\bibitem [{\citenamefont {Borovkov}(1984)}]{Borovkov1984}%
  \BibitemOpen
  \bibfield  {author} {\bibinfo {author} {\bibfnamefont {A.~A.}\ \bibnamefont
  {Borovkov}},\ }\href@noop {} {\emph {\bibinfo {title} {Mathematical
  Statistics. Parameter Estimation}}}\ (\bibinfo  {publisher} {Moscow: Nauka},\
  \bibinfo {year} {1984})\BibitemShut {NoStop}%
\bibitem [{\citenamefont {van Trees}(1968)}]{vantrees1968}%
  \BibitemOpen
  \bibfield  {author} {\bibinfo {author} {\bibfnamefont {H.~L.}\ \bibnamefont
  {van Trees}},\ }\href@noop {} {\emph {\bibinfo {title} {Detection, Estimation
  and Modulation Theory, Part 1}}}\ (\bibinfo  {publisher} {Wiley \& Sons},\
  \bibinfo {year} {1968})\BibitemShut {NoStop}%
\end{thebibliography}%

\clearpage
\onecolumngrid
\pagebreak
\widetext

\begin{center}
\textbf{\large Supplemental Materials: Overcoming Quantum Metrology Singularity through Sequential Measurements}

\vspace{0.25cm}

Yaoling Yang$^{1}$, Victor Montenegro$^{1,2}$, and Abolfazl Bayat$^{1,2}$

\vspace{0.25cm}

$^{1}${\small \em Institute of Fundamental and Frontier Sciences,\\ University of Electronic Science and Technology of China, Chengdu 610051, PR China}\\
$^{2}${\small \em Key Laboratory of Quantum Physics and Photonic Quantum Information, Ministry of Education,\\ University of Electronic Science and
Technology of China, Chengdu 611731, China}

\end{center}

\date{\today}
\setcounter{equation}{0}
\setcounter{figure}{0}
\setcounter{table}{0}
\setcounter{page}{1}
\makeatletter
\renewcommand{\theequation}{S\arabic{equation}}
\renewcommand{\thefigure}{S\arabic{figure}}

\section{I. Proof: Singularity of the Quantum Fisher Information Matrix Implies Singularity of Classical Fisher Information Matrix}

In this section, we show that the singularity of quantum Fisher information (QFI) matrix $\bm{\mathcal{Q}}(\bm{\lambda})$ implies that the classical Fisher information (CFI) matrix $\bm{\mathcal{F}}(\bm{\lambda})$ is also singular, where $\bm{\lambda}$ is the vector of unknown parameters, that is $\bm{\lambda}{=}(\lambda_1{,}\lambda_2{,}\ldots{,}\lambda_k)$. Let us recall the multi-parameter hierarchy of Fisher information matrices:
\begin{equation}
\bm{\mathcal{Q}}(\bm{\lambda}){\geq}\bm{\mathcal{F}}(\bm{\lambda})\ge0,\label{eq_Q_ge_C}
\end{equation}
The above implies $\bm{\mathcal{Q}}(\bm{\lambda})$, $\bm{\mathcal{F}}(\bm{\lambda})$, and $(\bm{\mathcal{Q}}(\bm{\lambda})-\bm{\mathcal{F}}(\bm{\lambda}))$ are all real positive semi-definite matrices. Therefore, $\bm{\mathcal{F}}(\bm{\lambda})$ can be diagonalized as $\bm{\mathcal{F}}(\bm{\lambda}){=} P\Lambda P^{-1}$, where $P$ is an orthogonal matrix and $\Lambda$ is a diagonal matrix with non-negative diagonal elements. 

Let us denote the square root of the above diagonal matrix as $\Lambda^{\frac{1}{2}}$. The square root of $\bm{\mathcal{F}}(\bm{\lambda})$, denoted by $S$, can be decomposed as $S{=}P\Lambda^{\frac{1}{2}}P^{-1}$ and is also a positive semi-definite matrix. One can check the following equality holds:
\begin{equation}
S^{-1} \bm{\mathcal{F}}(\bm{\lambda}) S^{-1}=(P\Lambda^{-\frac{1}{2}}P^{-1})(P\Lambda P^{-1})(P\Lambda^{-\frac{1}{2}}P^{-1})=I,
\end{equation}
where $I$ is an identity matrix.  

The determinant of $\bm{\mathcal{Q}}(\bm{\lambda})$ can be written as:
\begin{eqnarray}
    \det(\bm{\mathcal{Q}}(\bm{\lambda}))&=&\det(SS^{-1}\bm{\mathcal{Q}}(\bm{\lambda})S^{-1}S) \nonumber \\ 
    &=&\det(S^2)\det(S^{-1}\bm{\mathcal{Q}}(\bm{\lambda})S^{-1}) \nonumber \\ 
    &=&\det(\bm{\mathcal{F}}(\bm{\lambda}))\det(S^{-1}(\bm{\mathcal{Q}}(\bm{\lambda})-\bm{\mathcal{F}}(\bm{\lambda}))S^{-1}+I). \label{eq_inequality_det}
\end{eqnarray}

Recall the definition of real positive semi-definite matrices: for any vector $x\in \mathbb{R}^n$, $x^T\bm{\mathcal{F}}(\bm{\lambda})x\ge0$. By letting $S^{-1}v=x$, for any vector $v\in \mathbb{R}^n$, we have:
\begin{eqnarray}
   (v^T)S^{-1}(\bm{\mathcal{Q}}(\bm{\lambda})-\bm{\mathcal{F}}(\bm{\lambda}))S^{-1}v&=&(S^{-1}v)^T(\bm{\mathcal{Q}}(\bm{\lambda})-\bm{\mathcal{F}}(\bm{\lambda}))(S^{-1}v) \nonumber\\
   &=&x^T(\bm{\mathcal{Q}}(\bm{\lambda})-\bm{\mathcal{F}}(\bm{\lambda}))x\ge0.
\end{eqnarray}

Therefore, $S^{-1}(\bm{\mathcal{Q}}(\bm{\lambda})-\bm{\mathcal{F}}(\bm{\lambda}))S^{-1}$ is also a positive semi-definite matrix with the eigendecomposition: $S^{-1}(\bm{\mathcal{Q}}(\bm{\lambda})-\bm{\mathcal{F}}(\bm{\lambda}))S^{-1}{=}(P')\Lambda' (P')^{-1}$ where $P'$ and $\Lambda'$ are the orthogonal matrix and diagonal matrix, respectively. By adding an identity to the expression, one obtains:
\begin{equation}
S^{-1}(\bm{\mathcal{Q}}(\bm{\lambda})-\bm{\mathcal{F}}(\bm{\lambda}))S^{-1}+I=(P')(\Lambda')(P')^{-1}+(P')(P')^{-1}=(P')(\Lambda'+I)(P')^{-1}.
\end{equation}

Since all eigenvalues of the positive semi-definite matrix $S^{-1}(\bm{\mathcal{Q}}(\bm{\lambda})-\bm{\mathcal{F}}(\bm{\lambda}))S^{-1}$ are non-negative (the elements of $\Lambda'$ are non-negative), the eigenvalues of $S^{-1}(\bm{\mathcal{Q}}(\bm{\lambda})-\bm{\mathcal{F}}(\bm{\lambda}))S^{-1}+I$ must all be greater than or equal to 1 (the elements of $\Lambda'+I$ must all be greater than or equal to 1). Consequently, the determinant $\det(S^{-1}(\bm{\mathcal{Q}}(\bm{\lambda})-\bm{\mathcal{F}}(\bm{\lambda}))S^{-1}+I)\ge1$. According to Eq.~(\ref{eq_inequality_det}), the following inequality can be derived:
\begin{equation}    
\det(\bm{\mathcal{Q}}(\bm{\lambda}))\ge\det(\bm{\mathcal{F}}(\bm{\lambda})). 
\end{equation}

Therefore, when the determinant of the QFI matrix vanishes, i.e., $\det(\bm{\mathcal{Q}}(\bm{\lambda})){=}0$, it follows that $\det(\bm{\mathcal{F}}(\bm{\lambda})){=}0$. This means that the singularity of the QFI matrix implies the singularity of the CFI matrix, regardless of the measurement basis.

\section{II. Singularity of the classical Fisher information matrix}

\subsection{Singular classical Fisher information matrix with insufficient measurement outcomes: single qubit example}

In this section, we demonstrate that estimating two parameters $k{=}2$ with two measurements outcomes $m{=}2$ results in a singular classical Fisher information (CFI) matrix. To do so, let us consider a single qubit state $|\psi\rangle$ parameterized as:
\begin{equation}
|\psi(\theta,\phi)\rangle = \cos\frac{\theta}{2}\ket{\uparrow} + e^{i\phi}\sin\frac{\theta}{2}\ket{\downarrow},\label{eq_sm_psi}
\end{equation}
where $\theta$ and $\phi$ are the parameters to be estimated; $\ket{\uparrow}$ and $\ket{\downarrow}$ are the eigenstates of $\sigma_z$. Recall that the elements of the CFI matrix $\bm{\mathcal{F}}(\theta,\phi) = \begin{pmatrix}
\mathcal{F}_{\theta\theta} & \mathcal{F}_{\theta\phi} \\
\mathcal{F}_{\theta\phi} & \mathcal{F}_{\phi\phi}
\end{pmatrix}$ are defined as:
\begin{equation}
[\mathcal{F}(\theta,\phi)]_{ij}{=}\sum_{m} p(m|\theta,\phi) \left(\partial_i \ln p(m|\theta,\phi)\right) \left(\partial_j \ln p(m|\theta,\phi)\right),
\end{equation}
where $\partial_1{:=}\partial/\partial\theta$ and $\partial_2{:=}\partial/\partial\phi$, $p(m|\theta,\phi){=}|\bra{\Upsilon_m}\psi(\theta,\phi)\rangle|^2$, and $\ket{\psi}\bra{\psi}$ is the pure state shown in Eq.~\eqref{eq_sm_psi} encoding the multiple unknown parameters $(\theta,\phi)$. To evaluate the above CFI matrix, we consider a general measurement basis with projectors $\{\ket{\Upsilon_1}\bra{\Upsilon_1},\ket{\Upsilon_2}\bra{\Upsilon_2}\}$, where:
\begin{eqnarray}
     |\Upsilon_1\rangle&=&\cos\frac{\theta'}{2}\ket{\uparrow}{+}e^{i\phi'}\sin\frac{\theta'}{2}\ket{\downarrow},\\
      |\Upsilon_2\rangle&=&\sin\frac{\theta'}{2}\ket{\uparrow}{-}e^{i\phi'}\cos\frac{\theta'}{2}\ket{\downarrow}.
\end{eqnarray}
\noindent Explicit calculations of $p(m|\theta,\phi){=}|\bra{\Upsilon_m}\psi(\theta,\phi)\rangle|^2$ give the following matrix elements:
\begin{eqnarray}
    \mathcal{F}_{\theta\theta} &=& -\frac{(\sin \theta' \cos \theta \cos (\phi' -\phi )-\cos \theta' \sin \theta)^2}{(\sin \theta' \sin \theta \cos (\phi' -\phi )+\cos \theta' \cos \theta-1) (\sin \theta' \sin \theta \cos (\phi' -\phi )+\cos \theta' \cos
   \theta+1)},\\
    \mathcal{F}_{\phi\phi}& =& -\frac{\sin ^2(\phi' -\phi )}{(\cot \theta' \cot \theta+\cos (\phi' -\phi
   ))^2-\csc ^2\theta' \csc ^2\theta},\\
    \mathcal{F}_{\theta\phi}&=&\mathcal{F}_{\phi\theta}= \frac{\sin (\phi' -\phi ) (\cot \theta'-\cot \theta \cos (\phi' -\phi ))}{(\cot \theta' \cot \theta+\cos (\phi' -\phi ))^2-\csc ^2\theta' \csc ^2\theta}.
\end{eqnarray}
With the above matrix elements, the determinant of the CFI matrix is:
\begin{equation}
\det(\bm{\mathcal{F}}(\theta,\phi)) = \mathcal{F}_{\theta\theta}\mathcal{F}_{\phi\phi} - \mathcal{F}_{\theta\phi}^2 = 0.
\end{equation}
\noindent This result shows that the CFI matrix is universally singular for any measurement basis, regardless of the chosen arbitrary rotation angles $(\theta', \phi')$.

\red{For completeness, we now show that the above result also holds for a mixed qubit state. To this end, let us consider a mixture of an arbitrary single-qubit pure state $|\psi\rangle=\cos{\frac{\theta}{2}}\ket{\uparrow}+e^{i\phi}\sin{\frac{\theta}{2}}\ket{\downarrow}$ and the maximally mixed state $\frac{\mathcal{I}}{2}$:
\begin{equation}
    \rho = p|\psi\rangle \langle\psi| + (1-p)\frac{\mathcal{I}}{2},
\end{equation}
where $0 \leq p \leq 1$ is the classical mixing probability. The corresponding QFI matrix is
\begin{equation}
\mathcal{Q}(\theta,\phi) =\left(
\begin{array}{cc}
 p^2 & 0 \\
 0 & p^2 \sin ^2\theta  \\
\end{array}
\right).
\end{equation}
One can readily check that $\det[\mathcal{Q}(\theta,\phi)]>0$ for $p>0$ and $\theta\ne 0,\pi$, meaning the QFI matrix is non-singular.}

\red{By using the same general projective measurements from the main text: $\{|\Upsilon_1\rangle \langle\Upsilon_1|, |\Upsilon_2\rangle \langle\Upsilon_2|\}$, where
\begin{equation}
    |\Upsilon_1\rangle = \cos \frac{\theta'}{2} \ket{\uparrow} + e^{i\phi'} \sin \frac{\theta'}{2} \ket{\downarrow}, \hspace{0.5cm}
|\Upsilon_2\rangle = \sin \frac{\theta'}{2} \ket{\uparrow} - e^{i\phi'} \cos \frac{\theta'}{2} \ket{\downarrow},
\end{equation}
we can obtain the CFI matrix $\mathcal{F}$ with the following matrix elements:
\begin{align}
     \mathcal{F}_{\theta\theta} &= -\frac{p^2 (\sin \theta'  \cos \theta  \cos (\phi' -\phi )-\cos \theta'  \sin \theta )^2}{(p \sin \theta'  \sin \theta  \cos (\phi'
   -\phi )+p \cos \theta'  \cos \theta)^2 -1},  \\
    \mathcal{F}_{\phi\phi} &=-\frac{p^2 \sin ^2(\phi' -\phi )}{p^2 (\cot \theta'  \cot \theta +\cos (\phi' -\phi ))^2-\csc ^2\theta'
   \csc ^2\theta}, \\
   \mathcal{F}_{\theta\phi}  &=\mathcal{F}_{\phi\theta}=\frac{p^2
   \sin (\phi' -\phi ) (\cot \theta'-\cot \theta \cos (\phi' -\phi ))}{p^2 (\cot \theta' \cot \theta +\cos (\phi' -\phi ))^2-\csc
   ^2\theta'\csc ^2\theta}. 
\end{align}
With the above matrix elements, we can verify that for any $0 < p \leq 1$, we still have:
\begin{equation}
    \det[\mathcal{F}(\theta,\phi)]=\mathcal{F}_{\theta\theta}\mathcal{F}_{\phi\phi}-\mathcal{F}_{\theta\phi}^2 = 0.
\end{equation}
This demonstrates that the occurrence of a non-singular QFI matrix alongside a singular CFI matrix is not limited to pure states but persists across the entire Bloch sphere. More generally, as long as the number of measurement outcomes $m$ (which is $m = 2$ for our projective measurement) satisfies $m < k + 1$, where $k$ is the number of unknown parameters (here $k = 2$ for $\theta$ and $\phi$), the CFI matrix will be singular---regardless of whether the quantum state is pure or mixed.}

\subsection{Lifting the singularity of the classical Fisher information matrix with additional measurement outcomes}

We now demonstrate that by increasing the number of outcomes, one can resolve the singularity issue. Specifically, we consider a measurement strategy that probes the state $\ket{\psi(\theta, \phi)}$ shown in Eq.~\eqref{eq_sm_psi} using a set of POVMs in the $\sigma_z$ and $\sigma_x$ bases with equal weights.
\begin{eqnarray}
    |\Upsilon_1\rangle\langle\Upsilon_1|&=&\frac{1}{2}|0\rangle\langle0|,\\
    |\Upsilon_2\rangle\langle\Upsilon_2|&=&\frac{1}{2}|1\rangle\langle1|,\\
    |\Upsilon_3\rangle\langle\Upsilon_3|&=&\frac{1}{2}|+\rangle\langle+|,\\
    |\Upsilon_4\rangle\langle\Upsilon_4|&=&\frac{1}{2}|-\rangle\langle-|.
\end{eqnarray}
For the state $|\psi(\theta,\phi)\rangle$ shown in Eq.~\eqref{eq_sm_psi}, the above set of POVMs yield the following outcome probabilities:
\begin{eqnarray}
p(1|\theta,\phi) &=& \langle \psi(\theta,\phi) |\Upsilon_1\rangle\langle\Upsilon_1| \psi(\theta,\phi)\rangle = \frac{1}{2} \cos^2\left(\frac{\theta}{2}\right), \\
p(2|\theta,\phi) &=& \langle \psi(\theta,\phi) |\Upsilon_2\rangle\langle\Upsilon_2| \psi(\theta,\phi)\rangle =\frac{1}{2} \sin^2\left(\frac{\theta}{2}\right),  \\
p(3|\theta,\phi) &=& \langle \psi(\theta,\phi) |\Upsilon_3\rangle\langle\Upsilon_3| \psi(\theta,\phi)\rangle = \frac{1}{4} (1 + \cos\phi \sin\theta), \\
p(4|\theta,\phi) &=& \langle \psi(\theta,\phi) |\Upsilon_4\rangle\langle\Upsilon_4| \psi(\theta,\phi)\rangle = \frac{1}{4} (1 - \cos\phi \sin\theta).
\end{eqnarray}
With the above probability distributions, it is now straightforward to evaluate the CFI matrix as:
\begin{equation}
    \bm{\mathcal{F}}(\theta,\phi) = \begin{pmatrix}
\mathcal{F}_{\theta\theta} & \mathcal{F}_{\theta\phi} \\
\mathcal{F}_{\theta\phi} & \mathcal{F}_{\phi\phi}
\end{pmatrix} = \begin{pmatrix}
        \frac{-\cos^2\phi\cos(2\theta) - 1}{2(\cos^2\phi\sin^2\theta - 1)} & 
        \frac{\sin\phi\cos\phi\sin\theta\cos\theta}{2(\cos^2(\phi)\sin^2(\theta) - 1)} \\[6pt]
        \frac{\sin\phi\cos\phi\sin\theta\cos\theta}{2(\cos^2(\phi)\sin^2(\theta) - 1)} & 
        \frac{1}{2\csc^2(\phi)\csc^2(\theta) - 2\cot^2(\phi)}
    \end{pmatrix},
\end{equation}
with determinant:
\begin{equation}
    \det(\bm{\mathcal{F}}(\theta,\phi))=\frac{1}{4 \csc^2(\phi) \csc^2(\theta)-4 \cot^2(\phi)}.
\end{equation}
Crucially, this determinant is generally non-zero, with $\det(\bm{\mathcal{F}}(\theta, \phi))$ vanishing only at very specific values, namely on the XZ-plane (with the choice of POVMS above). Additionally, note that the single-qubit example is the simplest case, and implementing multiple POVMs may not always be feasible in practice. This demonstrates the benefits of our multi-parameter sequential measurement scheme, where a local and fixed measurement is used throughout the entire sensing protocol.

This result shows that incrementing the number of measurement outcomes to a four-outcome measurement scheme allows for the simultaneous estimation of both parameters, unlike the two-outcome case discussed in the previous section. 

\begin{figure}[b]
\includegraphics[width=0.85\linewidth]{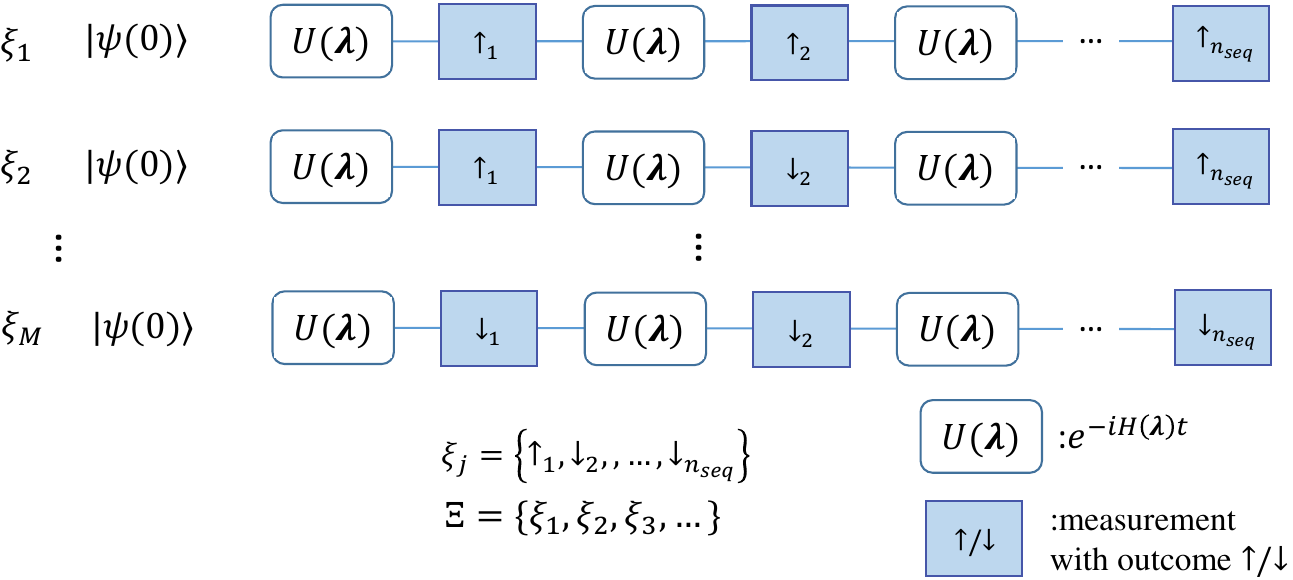}  
\caption{A schematic of the sequential measurement sensing protocol~\cite{montenegro2022sequential}. A quantum trajectory $\xi_j$ is constructed by starting with an initial state $\ket{\psi(0)}$, which undergoes a sequence of evolution steps governed by the operator $U(\bm{\lambda})$ to encode the unknown parameters $\bm{\lambda}$. Each evolution step is followed by a measurement, with outcomes $\uparrow_i$ or $\downarrow_i$, resulting in a sequence of $n_\mathrm{seq}$ outcomes, for instance $\xi_j = \{\uparrow_1, \downarrow_2, \dots, \downarrow_{n_\mathrm{seq}}\}$. While the evolution is assumed to be unitary, it can be generalized to non-unitary dynamics~\cite{yang2024sequential}. After completing $n_\mathrm{seq}$ measurements for a single trajectory, the quantum probe is reset to its initial state, and the process is repeated to generate a new trajectory.}\label{fig_SM_schematic} 
\end{figure}

\section{III. Bayesian posterior}

In sequential measurements sensing, the observed data is a collection of $\pmb{\xi}_j$ quantum trajectories given by:
\begin{equation}
\pmb{\Xi}{=}\{\pmb{\xi}_1,\pmb{\xi}_2,\cdots,\pmb{\xi}_M\},
\end{equation}
where $\pmb{\Xi}$ is the array of $M$ collected trajectories, each $\pmb{\xi}_j$ consisting of $n_\mathrm{seq}$ measurement outcomes obtained from probing a local spin with a fixed measurement basis. A schematic of the sequential measurement sensing protocol is shown in Fig.~\ref{fig_SM_schematic}. The likelihood function $P(\text{data}|\bm{\lambda})$ in Eq.~\eqref{eq_Bayes_rule} is defined as~\cite{montenegro2022sequential}:
\begin{equation}
P(\text{data}=\bm{\Xi}|\bm{\lambda})=\frac{M!}{s_1!s_2!\cdots s_{2^{n_\mathrm{seq}}}!}\prod_{j=1}^{2^{n_\mathrm{seq}}}\left[p(\pmb{\xi}_j|\bm{\lambda})\right]^{s_j},
\end{equation}
where $p(\pmb{\xi}_j|\bm{\lambda})$ is the conditional probability for quantum trajectory $\pmb{\xi}_j$ assuming $\bm{\lambda}$; $s_1,{\cdots},s_{2^{n_\mathrm{seq}}}$ represent the number of times that sequences from $(\uparrow_1,\uparrow_2,\ldots,\uparrow_{n_\mathrm{seq}})$ to $(\downarrow_1,\downarrow_2,\ldots,\downarrow_{n_\mathrm{seq}})$ appears across the entire data set of $M$ samples. Here, $\uparrow_j$ and $\downarrow_j$ are the $\sigma_z$ eigenvalues for the $j$th measurement instance. Note that $p(\pmb{\xi}_j|\bm{\lambda})$ only requires the quantum statistical model (i.e., the state $\rho(\bm{\lambda})$) and the choice of POVM for its evaluation, which must be classically simulated for all possible sequences ($2^{n_\mathrm{seq}}$ in total) over a relevant range of $\bm{\lambda}$. Up to this point, we have a set of probability distributions for a range of multiple parameters $\bm{\lambda}$ and quantum trajectories $\pmb{\xi}_j$. The likelihood function then takes the observed data (either from experiments or randomly simulated) and assigns which parameters best fit the observed data. The posterior is then straightforwardly evaluated as:
\begin{equation}
P(\bm{\lambda}|\text{data}=\bm{\Xi}) = \frac{P(\text{data}=\bm{\Xi}|\bm{\lambda})P(\bm{\lambda})}{P(\text{data}=\bm{\Xi})},
\end{equation}
as shown in Eq.~\eqref{eq_Bayes_rule} in the main text. For simplicity, the prior $P(\bm{\lambda})$ can be assumed to be uniformly distributed, while $P(\text{data} = \bm{\Xi})$ serves as a normalization constant ensuring that the posterior integrates to unity over the entire hypervolume $d^k\bm{B}$.

\section{IV. Bayesian estimation for the Jaynes-Cummings probe}

\begin{figure}
\includegraphics[width=0.95\linewidth]{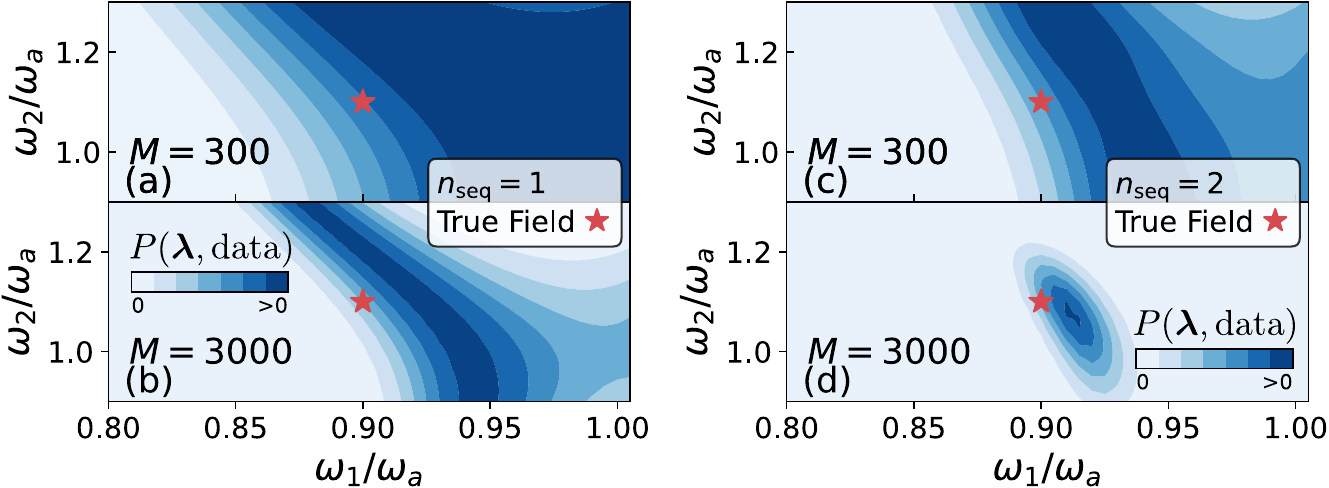}  
\caption{Posterior functions for the light-matter probe as functions of $\omega_1$ and $\omega_2$ for different $n_\mathrm{seq}$ and $M$. Left (right) column corresponds to $n_\mathrm{seq}{=}1$ ($n_\mathrm{seq}{=}2$). The true parameters are set to ($\omega_1{,}\omega_2){=}(0.9{,}1.1)\omega_a$.} \label{fig_SM_posterior_JC} 
\end{figure}

In this section, we perform a Bayesian analysis of the light-matter probe described in Eq.~\eqref{eq_JC_model}. First, we assume that only two parameters are unknown: the frequencies of the two-level atoms, $\omega_1$ and $\omega_2$. Following the Bayesian procedure outlined above, we construct the posterior function from the probability distributions obtained by measuring a single two-level atom sequentially at time intervals $\omega_a t = 2\pi$ in the $\sigma_z$ basis. In Fig.~\ref{fig_SM_posterior_JC}, we plot the posterior as function of $\omega_1$ and $\omega_2$ for different numbers of sequential measurements $n_\mathrm{seq}$ and sample sizes $M$.  

Figs.~\ref{fig_SM_posterior_JC}(a)-(b) depict the posterior function for $n_\mathrm{seq}{=}1$ with two different sample sizes $M$. As shown in the figures, the posterior function fails to assign a unique set of values to $\omega_1$ and $\omega_2$. This occurs because the condition for the invertibility of the CFI matrix, namely $2^{n_\mathrm{seq}}{\geq}k{+}1$, where $k$ is the number of unknown parameters (in this case, $k=2$), is not satisfied. Consequently, the posterior function in Figs.~\ref{fig_SM_posterior_JC}(a)-(b) reflects the challenge of estimating a set of unknown parameters when the CFI matrix is singular.  

In contrast, Figs.~\ref{fig_SM_posterior_JC}(c)-(d) show the posterior function for $n_\mathrm{seq}{=}2$ with two different sample sizes $M$. Here, the posterior function successfully assigns a unique set of values to $\omega_1$ and $\omega_2$, as the condition for the invertibility of the CFI matrix ($2^2{\geq}3$) is now satisfied. This demonstrates that simultaneous estimation of $\omega_1$ and $\omega_2$ becomes feasible when the number of sequential measurements is increased to $n_\mathrm{seq}{=}2$. 

To further assess the estimability of $\bm{\lambda}$, we analyze the covariance matrix of the estimator $\hat{\bm{\lambda}}$, as described in Eq.~\eqref{eq_trace_QCRB} in the main text. The estimated parameter is determined by identifying the coordinate that maximizes the posterior function for a given $M$:  
\begin{equation}
    \boldsymbol{B}_{\mathrm{est},i} = \underset{\bm{B}}{\arg\max} \, P(\boldsymbol{B}|\text{data}),
\end{equation}  
see the ``Bayesian estimation" section in the main text for details.  

\begin{figure}
\includegraphics[width=0.8\linewidth]{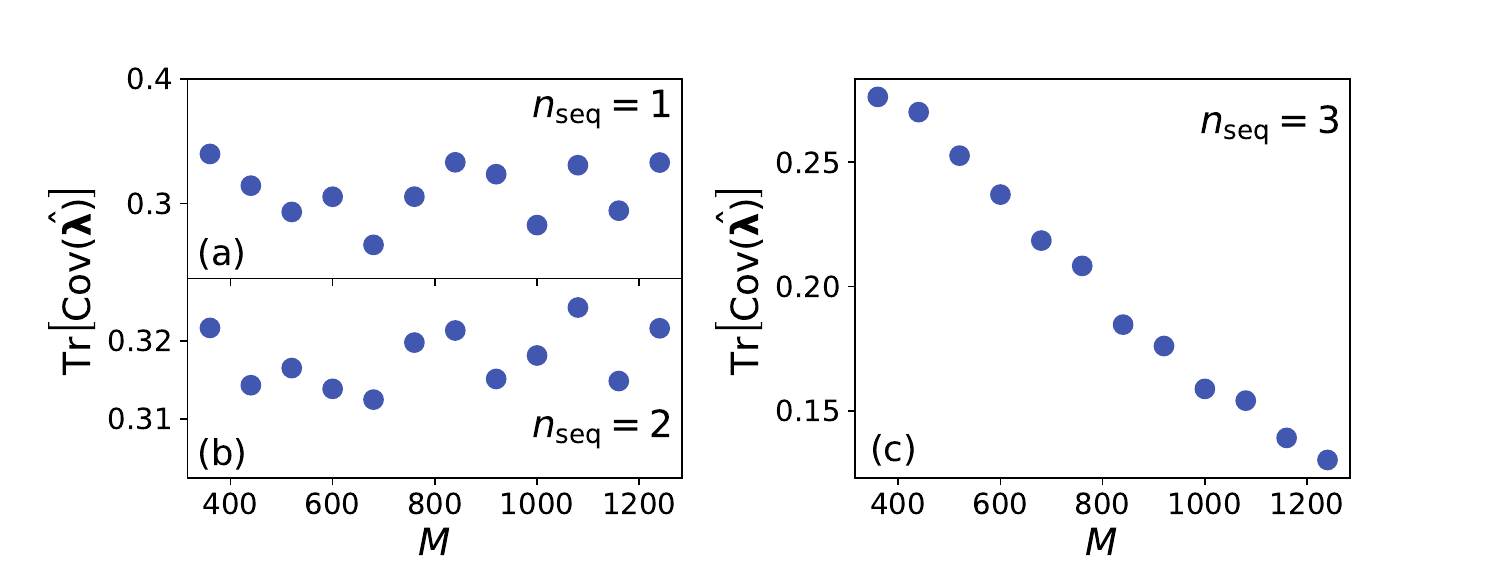}  
\caption{Trace of covariance matrix $\text{Tr}(\text{Cov}[\hat{\boldsymbol{\lambda}}])$ using a Bayesian estimator as a function of $M$. We encode $k{=}4$ unknown parameters $\bm{\lambda}{=}(\omega_1{,}\omega_2{,}J_1{,}J_2)$. (a) $n_\text{seq}{=}1$, (b) $n_\text{seq}{=}2$, and (c) $n_\text{seq}{=} 3$. The parameters to be estimated are set to $\bm{\lambda}{=}(0.8{,}0.9{,}0.2{,}0.3)\omega_a$.} \label{fig_SM_trace_COV} 
\end{figure}

In Fig.~\ref{fig_SM_trace_COV}, we plot the trace of the covariance matrix $\text{Tr}(\text{Cov}[\hat{\bm{\lambda}}])$ as a function of $M$ for different values of $n_\mathrm{seq}$, encoding $k{=}4$ parameters $(\omega_1, \omega_2, J_1, J_2)$. Figs.~\ref{fig_SM_trace_COV}(a)-(b) show that when $n_\mathrm{seq}{<}3$, the estimation of four parameters is not possible, as $\text{Tr}(\text{Cov}[\hat{\bm{\lambda}}])$ remains nearly constant with increasing $M$, indicating that the CFI matrix is singular. This occurs because the invertibility condition of the CFI matrix, $2^{n_\mathrm{seq}}{\geq}k{+}1$ (i.e., $2^{n_\mathrm{seq}}{\geq}5$), is not satisfied for $n_\mathrm{seq}{<}3$.  

In contrast, Fig.~\ref{fig_SM_trace_COV}(c) shows that for $n_\mathrm{seq}{=}3$, $\text{Tr}(\text{Cov}[\hat{\bm{\lambda}}])$ decreases noticeably with increasing $M$, consistent with the condition $2^{n_\mathrm{seq}}{\geq}k{+}1$ being met. This result highlights the effectiveness of our sequential measurements protocol in simultaneously estimating multiple parameters.

\section{V. Protocol Optimization}

\red{The sequential measurement sensing protocol in the main text operates with minimal control over the system, namely uses fixed measurements in the $\sigma_z$ basis, and assumes a fixed evolution time interval $J\tau = N$ for simplicity. In this section, we investigate how performance changes when separately optimizing two parameters at each sequential step: (1) the measurement basis and (2) the evolution time interval. For clarity, we focus on the spin chain Hamiltonian introduced in Eq.~\eqref{eq_heisenberg_model} of the main text, assuming a system size of $N=4$ and encoding $k=2$ parameters.}

\subsection*{\red{Optimization of measurement basis}}

\red{To optimize the measurement basis, we consider projective measurements with operators ${\mathcal{I}\otimes|\Upsilon_1\rangle\langle\Upsilon_1|, \mathcal{I}\otimes|\Upsilon_2\rangle\langle\Upsilon_2|}$, where:
\begin{eqnarray}
     |\Upsilon_1\rangle&=&\cos\frac{\theta'}{2}\ket{\uparrow}{+}e^{i\phi'}\sin\frac{\theta'}{2}\ket{\downarrow},\\
      |\Upsilon_2\rangle&=&\sin\frac{\theta'}{2}\ket{\uparrow}{-}e^{i\phi'}\cos\frac{\theta'}{2}\ket{\downarrow}.
\end{eqnarray}
with $0 \leq \phi' \leq 2\pi$, $0 \leq \theta' \leq \pi$, and $\mathcal{I}$ being an identity operator. These measurement operators reduce to the standard $\sigma_z$ basis when $\theta'=0$ and $\phi'=0$.}

\red{For the sake of simplicity, we assume the total number of sequential measurements to be $n_{\mathrm{seq}} = 4$. Two optimization methods are implemented:
\begin{itemize}
    \item Grid search, and
    \item Greedy search.
\end{itemize}}

\red{In the grid search, we identify optimized measurement bases by systematically exploring the parameter space of $\theta'$ and $\phi'$ at each measurement step. Due to computational resource constraints, we uniformly sample seven values for each parameter and evaluate all possible combinations across the four sequential measurements.}

\red{The greedy approach proceeds sequentially, optimizing one measurement at a time starting from the second measurement. The first measurement is fixed at $\theta' = 0$ and $\phi' = 0$, as a single measurement leads to a singular CFI matrix. For each subsequent step, we perform a grid search over $\theta'$ and $\phi'$ (sampling 7 values for each parameter), while keeping the previously optimized parameters fixed. This process continues until all $n_{\mathrm{seq}}$ measurement steps are determined.}

\begin{figure}
\centering
\includegraphics[width=\linewidth]{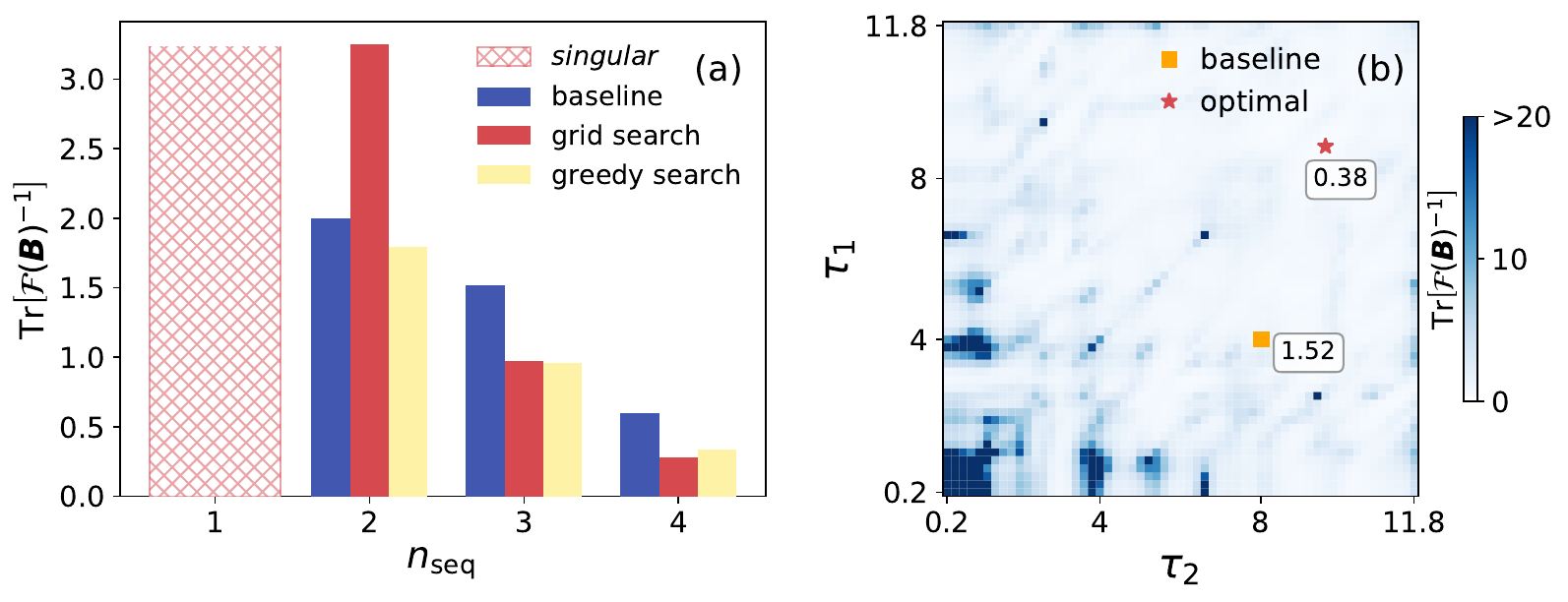} 
\caption{\red{(a) The trace of the inverse of the CFI matrix $\mathrm{Tr}[\mathcal{F}(\bm{B})^{-1}]$ as a function of the number of sequential measurements $n_{\mathrm{seq}}$ for baseline (no optimization $\sigma_z$ measurement basis) (blue bar), grid search (red bar), and greedy search (yellow bar) optimizations. The lower values represent better sensing precision, while the cross-hatched region corresponds to a singular CFI matrix. (b) Heatmap of $\mathrm{Tr}[\mathcal{F}(\bm{B})^{-1}]$ versus measurement timings $\tau_1$ and $\tau_2$ for $n_{\mathrm{seq}}=3$. The orange square shows baseline uniform timing (1.52), while the red star indicates optimal non-uniform timing (0.38). Lower values represent better sensing precision.} } \label{fig_SM_optimization} 
\end{figure}

\red{To evaluate performance for $n_{\mathrm{seq}}=4$, we use the same metric as in the main text, namely $\mathrm{Tr}[\mathcal{F}(\bm{B})^{-1}]$, as a function of the number of sequential measurements $n_{\mathrm{seq}}$. In Fig.~\ref{fig_SM_optimization}(a), we compare sensing performance between our original protocol (baseline, blue bar) using the $\sigma_z$ basis and the optimized protocols (red for grid search and yellow for greedy search). Both optimized protocols show improved sensing precision at $n_{\mathrm{seq}}=4$. Interestingly, the grid search degrades performance at $n_{\mathrm{seq}} = 2$, but outperforms both the baseline and the greedy search method at $n_{\mathrm{seq}} = 4$. The greedy search method offers lower computational cost while maintaining good sensing precision compared to the baseline, although it performs slightly worse than the grid search at $n_{\mathrm{seq}} = 4$. These results demonstrate that significant improvements in multi-parameter estimation can be achieved by selecting an appropriate measurement basis in sequential measurement sensing protocols.}

\subsection*{\red{Optimization of evolution time intervals}} 

\red{Another optimization strategy is to vary the time intervals between measurements instead of using fixed intervals. To clarify this idea, consider a scenario with $n_{\mathrm{seq}} = 3$ sequential measurements on a spin chain of length $N = 4$, and a fixed total evolution time given by $JT = n_{\mathrm{seq}} N$. We assume that all measurements are performed in the $\sigma_z$ basis.}

\red{In the original unoptimized (baseline) protocol, consecutive measurements are evenly spaced in time: the system evolves for a fixed interval $\tau$ between each measurement, with $\tau_1 = \tau_2 = \tau_3 = \tau$, such that the total time satisfies $\tau_1 + \tau_2 + \tau_3 = T$. This choice corresponds to $J \tau = N$, and is motivated both by simplicity and by the Lieb–Robinson bound. However, instead of keeping all time intervals equal, we can improve the precision by optimizing the timing of the first two measurements. Specifically, we denote the evolution time before the first measurement as $\tau_1$, and the time between the first and second measurements as $\tau_2$. The third interval $\tau_3$ is then fixed by the constraint $\tau_1 + \tau_2 + \tau_3 = T$, so that $\tau_3 = T - \tau_1 - \tau_2$.}

\red{In Fig.~\ref{fig_SM_optimization}(b), we present a heatmap of $\mathrm{Tr}[\mathcal{F}(\bm{B})^{-1}]$ as functions of $\tau_1$ and $\tau_2$ for $n_\mathrm{seq}=3$ and $N=4$. The square marker represents the performance of the original protocol, where all evolution times are equal and set to $J\tau = N$. This gives a precision of $\mathrm{Tr}[\mathcal{F}(\bm{B})^{-1}] \approx 1.52$. In contrast, the star marker corresponds to the optimized protocol with non-uniform measurement timings. This optimized strategy significantly improves sensing precision, reducing the value to $\mathrm{Tr}[\mathcal{F}(\bm{B})^{-1}] \approx 0.38$. This result demonstrates that allowing non-uniform evolution times between measurements can lead to a substantial enhancement in performance. Finally, while this work focuses on optimizing the timing of measurements, other strategies could also be considered. These include jointly optimizing both the measurement basis and the evolution times, or employing learning algorithms tailored to specific sensing resources. Although our results clearly demonstrate that optimizing measurement timing leads to a significant improvement in sensing precision, a comprehensive analysis of these broader optimization approaches lies beyond the scope of this work.}

\end{document}